\documentclass[11pt, a4paper]{article}
\usepackage{setspace}
\usepackage{pdflscape}
\usepackage{float}
\usepackage{placeins}
\usepackage[T1]{fontenc}
\usepackage{newtxtext,newtxmath}
\usepackage[numbers,sort&compress]{natbib}
\usepackage{tabularx}
\usepackage{array}
\usepackage{ragged2e}

\usepackage{amsmath, amssymb, amsfonts, mathtools}
\usepackage[a4paper, margin=1in]{geometry}         
\usepackage{graphicx}
\usepackage{xcolor}
\usepackage{hyperref} 
\usepackage{booktabs} 
\usepackage{siunitx}
\usepackage{textgreek}

\newcommand{\xs}{x_s}

\newcommand{\deltah}{\delta_{\mathrm{h}}}

\usepackage{subcaption}
\usepackage{titling}
\usepackage{authblk}
\usepackage{pgfplots}
\pgfplotsset{compat=1.17}
\usetikzlibrary{math}

\hypersetup{
  colorlinks=true,
  linkcolor=blue!50!black,
  urlcolor=blue!50!black,
  citecolor=blue!50!black
}
\providecommand{\keywords}[1]{\par\vspace{0.5em}\noindent\textbf{Keywords:} #1\par\vspace{0.5em}}

\title{\Large \bf
Shock-Centered Low-Rank Structure and Neural-Operator Representation of Rarefied Micro-Nozzle Flows}
\author[1,*]{Ehsan Roohi\thanks{Lecturer}}
\author[2]{Amirmehran Mahdavi\thanks{Assistant professor}}

\affil[1]{Mechanical and Industrial Engineering, University of Massachusetts Amherst, 160 Governors Dr., Amherst, 01003}
\affil[2]{Department of Mechanical Engineering, Hakim Sabzevari University, Sabzevar, Iran}
\affil[*]{Corresponding author: \href{mailto:roohie@umass.edu}{roohie@umass.edu}}

\date{\today}
\begin{document}
\doublespacing
\maketitle

\begin{abstract}
We examine the structure of Direct Simulation Monte Carlo (DSMC)-resolved
internal compression layers in rarefied micro-nozzle flows and show that their
apparent parametric complexity is largely a registration and finite-thickness
scaling effect. A density-gradient diagnostic identifies the compression-layer
station \(x_s\), while a jump-based thickness
\(\delta_j=\Delta\rho/\max|\partial\rho/\partial x|\) defines a shock-centered
coordinate \(\xi_j=(x-x_s)/\delta_j\). In physical coordinates, the leading
proper orthogonal decomposition (POD) mode of the centerline density profiles
captures only \(83.33\%\) of the fluctuation energy, whereas the jump-scaled
coordinate increases this value to \(98.33\%\). A two-dimensional
shock-window POD further confirms that this compactness is not a centerline
artifact: in the registered \((\xi_j,\eta)\) frame, the first density mode
captures \(94.98\%\) and the first two modes capture \(99.05\%\) of the
fluctuation energy. The same region is identified by density-gradient and
gradient-length Knudsen-number diagnostics, linking the reduced representation to localized short-gradient-length
rarefaction rather than to shock motion alone. We then use
this structure as an inductive bias in a shock-aligned Fusion--Deep Operator
Network (DeepONet) surrogate for density, velocity components, temperature,
Mach number, and pressure. For held-out back-pressure cases, density,
temperature, and pressure errors remain below \(6.8\%\), \(4.3\%\), and
\(6.8\%\), respectively, and the hardest case reduces the shock-window mean
error from \(9.75\%\)--\(22.27\%\) for standard baselines to \(4.51\%\). The
results show that improved prediction follows from the reduced shock-centered
structure of the DSMC fields rather than from network capacity alone.
\end{abstract}



\section{Introduction}
\label{sec:introduction}

Micro-nozzles are key components in modern aerospace and micro-scale technologies, providing compact thrust for satellite attitude control, station-keeping, micro- and nano-spacecraft propulsion, and related micro-electro-mechanical systems (MEMS)-scale devices  \cite{alexeenko2002numerical, saadati2015detailed,jin2024numerical,mutis2024numerical, nishii2025numerical, sabouri2025propulsive,kumar2025plume}. Beyond space-propulsion applications, they are also relevant to vacuum-generation technologies and materials-processing systems where controlled high-speed gas jets are required. In a converging--diverging micro-nozzle, gas accelerates from subsonic conditions to sonic speed near the throat and then expands into the diverging section. At the microscale, however, this apparently classical process becomes strongly affected by rarefaction, non-equilibrium transport, wall interactions, and shock-wave formation.

The central difficulty is that the flow may span multiple rarefaction regimes within the same device. Because the characteristic dimensions are small and the pressure gradients are large, the local Knudsen number, $Kn$, can vary substantially along the nozzle \cite{bird1994molecular}. The flow may begin near the continuum regime, pass through slip-flow conditions, and locally approach transitional behavior in the downstream or shock-affected regions \cite{karniadakis2005microflows}. Under these conditions, the assumptions underlying the Navier--Stokes--Fourier equations become increasingly questionable \cite{karniadakis2005microflows}. The challenge is further intensified by internal shock waves whose location and strength depend sensitively on the imposed back pressure \cite{darbandi2011study}. These shocks introduce steep gradients in velocity, pressure, density, and temperature, and represent localized regions of strong thermodynamic non-equilibrium. Recent studies have also emphasized that nonequilibrium shock structure and
continuum breakdown depend sensitively on relaxation physics and local
breakdown indicators~\cite{swaminathan2016gce_breakdown,zeng2025bulk_viscosity_shock}.
Consequently, micro-nozzle prediction is not simply a problem with a single
global rarefaction level, but a multi-regime, spatially heterogeneous
kinetic-flow problem. 

For such rarefied and non-equilibrium conditions, the Direct Simulation Monte Carlo (DSMC) method remains one of the most reliable high-fidelity approaches. DSMC, originally developed by Bird \cite{bird1994molecular}, solves the gas dynamics statistically by tracking representative particles and modeling molecular motion and collisions. It has been widely applied to rarefied flows and micro-nozzle problems, including studies of back pressure, geometry, gas species, and shock structures \cite{stefanov2019basic,ivanov1999numerical,xie2007micronozzle,darbandi2011study, saadati2015detailed,la2011hybrid, mahdavi2020novel}. Nevertheless, DSMC is computationally expensive, especially in slip and early transitional regimes where many particles and long sampling times are required \cite{darbandi2011study}. This cost becomes a major obstacle for design-oriented tasks such as parametric sweeps, uncertainty quantification, optimization, and repeated evaluation over back pressure or geometric variables \cite{Prakash2025BCMLSupersonic,Prakash2026BCMLReactiveRarefied}.

Several reduced-cost strategies have therefore been explored. In our previous work on a related micro-nozzle configuration, a hybrid DSMC--Fokker--Planck approach was used to reduce computational cost by applying DSMC mainly in strongly non-equilibrium regions and a faster Fokker--Planck solver elsewhere \cite{mahdavi2020novel}. Although this hybrid strategy reduced cost, it also highlighted an important limitation: some shock-related flow features were not captured with sufficient fidelity \cite{mahdavi2020novel}. This motivates the development of surrogate models that retain the accuracy of DSMC data while providing the speed required for repeated engineering evaluation.

Deep-learning-based and physics-aware surrogate models have recently become a
promising route for accelerating high-fidelity flow prediction
\cite{tatsios2025dsmc,nair2026physics_ml_closures,roohi2026pof_rgd_nn,he2026mdcnn,Prakash2025BCMLSupersonic,Prakash2026BCMLReactiveRarefied}. Purely data-driven networks, such as fully connected neural networks or convolutional architectures, can provide substantial acceleration but may suffer from limited physical consistency, poor extrapolation, and smoothing of localized high-gradient structures \cite{ma2026locally}. Physics-Informed Neural Networks (PINNs) address some of these issues by embedding governing-equation residuals into the training loss \cite{raissi2019physics}. This can improve physical plausibility and regularization \cite{ma2026locally}, but standard PINNs are usually designed to solve one parameter instance at a time and are therefore less convenient for many-query parametric studies \cite{roohi2025analysis}.

Operator-learning methods provide a more suitable framework for parametric prediction. In particular, DeepONet learns a mapping from input parameters or functions to output solution fields \cite{lu2021learning}. Its branch network encodes the operating condition, while its trunk network represents the spatial dependence of the solution. The present micro-nozzle problem naturally has this form: for a given back pressure, the goal is to learn the operator
\[
f_{\theta}:\;(P_{\mathrm{back}},x,y)\mapsto
(\rho,U,V,T,\mathrm{Mach},P).
\]
However, neural operators still face a major difficulty when the target field contains moving shocks or localized compression layers. Standard losses often emphasize global accuracy and may smear sharp features, especially when a shock shifts with the input parameter.

Several strategies have been proposed to improve shock or high-gradient reconstruction. Ma et al. developed a Locally Enhanced PINN for aero-engine nozzles, using a specialized local network to improve near-wall pressure and temperature prediction \cite{ma2026locally}. Our previous work used zonal loss weighting to emphasize physically important flow regions in rarefied micro-step flows \cite{roohi2025analysis}. More recently, Peyvan et al. introduced Fusion--DeepONet for geometry-dependent hypersonic and supersonic flows, in which branch and trunk representations are coupled through multi-scale multiplicative fusion rather than only a final inner product \cite{PEYVAN2026114432}. Related coordinate-alignment ideas also appear in flexDeepONet, where a learned preprocessing transformation aligns transported structures before operator learning \cite{venturi2023svd}, and in nonlinear-manifold operator approaches such as NOMAD, which replace purely linear reconstruction by nonlinear decoding \cite{seidman2022nomad}. Related ideas also appear in the reduced-order modeling literature for high-speed flows with moving discontinuities. Classical POD-based reduced-order models become inefficient when the dominant flow variation is the motion of a shock rather than a smooth amplitude change. For high-speed flows with moving shocks, domain-decomposition and transported-snapshot strategies have therefore been introduced to separate or register the moving discontinuity before constructing a compact basis \cite{lucia2003movingShocks,nair2019transportedSnapshots}. Similarly, shifted POD formulations have shown that transport-dominated structures can become substantially more compact when the data are represented in a co-moving frame \cite{reiss2018shiftedPOD}. These studies motivate the present shock-centered analysis. The present work differs, however, in that the registration scale is extracted from DSMC density-gradient diagnostics of a finite-thickness rarefied compression layer and is then embedded directly into the trunk features of a neural operator. Thus, the novelty is not the use of registration alone, but the identification
of a finite-thickness kinetic compression layer whose DSMC-resolved variability
collapses under a jump-based scale tied to the local mean free path and
gradient-length rarefaction diagnostics. This provides a physical basis for
the shock-aligned neural-operator features, rather than treating alignment as
a purely numerical preprocessing step.

In parallel, discontinuity-aware learning strategies have been proposed for shock-dominated or non-smooth fields. These include Kolmogorov--Arnold network (KAN)-based discontinuity-aware PINNs \cite{lei2025discontinuity}, adaptive-mesh transformer models \cite{xu2025amr}, and the recently proposed \(\phi\)-DeepONet framework for operator learning with discontinuous inputs or outputs \cite{roy2026phideeponet}. The latter is particularly relevant because it modifies the DeepONet representation to account for interface-induced discontinuities through a domain/interface-aware trunk embedding. In contrast, the present work does not impose interface-informed residual losses or explicit domain-decomposition labels; instead, it introduces a DSMC-derived shock-distance coordinate and multi-scale shock envelopes as low-cost trunk features for a moving internal compression layer.

Surrogate modeling of rarefied plume and nozzle-related flows has also advanced rapidly. For example, He et al. developed a multi-decoder ConvNeXt-form network for DSMC-based vacuum plume prediction in variable-thrust engines \cite{he2026mdcnn}. Their model combines carefully designed input features with a multi-decoder architecture and shows improved prediction accuracy relative to a vanilla U-Net. This recent work is closely related in motivation because it addresses the high cost of DSMC-based aerospace flow prediction. The present study differs in focus and mechanism: rather than reconstructing vacuum plume fields with a convolutional image-to-image architecture, we develop a shock-aligned operator-learning framework for parameter-dependent rarefied micro-nozzle flows, where the dominant challenge is reconstructing flow fields containing a moving internal compression layer.

In this work, we first examine whether the DSMC-resolved internal compression layer admits a shock-centered reduced representation. We then use the resulting low-rank structure as an inductive bias in a Fusion--DeepONet operator model. The proposed method is physics-motivated rather than residual-constrained:
it does not enforce governing-equation residuals, conservation constraints,
or Rankine--Hugoniot jump conditions in the loss. Instead, physical structure
is introduced through training-data-derived shock-aligned features and
shock/gradient-emphasized sample weighting. The trunk input is augmented by shock-aligned features, including a signed distance to an estimated shock station, a smooth pre-/post-shock indicator, and multi-scale Gaussian envelopes around the shock region. These features act as an interpretable inductive bias, analogous in spirit to signed-distance or wall-distance features used in geometry-aware surrogates. They reduce the burden on the neural operator to discover a moving discontinuity directly from Cartesian coordinates alone.

The proposed framework combines three elements. First, a Fusion--DeepONet architecture is used to represent the parameter-to-field mapping. Second, shock-aligned trunk features provide a localized coordinate representation of the moving compression layer. Third, a two-phase curriculum weighting strategy emphasizes high-gradient regions during training. 

The main contributions of this work are as follows:
\begin{itemize}
\item A density-gradient and gradient-length Knudsen-number analysis showing
that the internal compression layer is a localized finite-thickness kinetic
structure rather than a discontinuous surface.

\item A shock-centered reduced representation of the DSMC-resolved compression
layer, showing that jump-scaled registration increases the leading centerline
POD energy from \(83.33\%\) to \(98.33\%\), reduces the number of modes needed
for 99\% energy, and produces a compact two-dimensional shock-window density
representation with \(99.05\%\) energy captured by the first two modes.

\item A kinetic interpretation of the reduced coordinate through the layer
thickness \(\delta_j\), local mean free path \(\lambda_s\), and gradient-length
rarefaction diagnostics.

\item A shock-aligned neural-operator representation that embeds this reduced
structure through signed distance, smooth pre-/post-layer indicators, and
Gaussian envelopes.

\item Validation on held-out back-pressure and throat-location cases,
demonstrating that the improved shock-window accuracy follows from the reduced
shock-centered structure rather than from network capacity alone.
\end{itemize}

Before applying the model to the micro-nozzle problem, we also use the viscous 1D Burgers' equation as a controlled sanity check for shock-like gradient reconstruction. This benchmark is reported in the Appendix A. The remainder of the paper is organized as follows. Section~\ref{sec:setup} describes the DSMC data generation, physical configuration, and train/test split. Section~\ref{sec:reduced_structure} analyzes the DSMC compression layer using density-gradient, gradient-length Knudsen-number, shock-centered collapse, and POD diagnostics. Section~\ref{sec:shock_features} introduces the shock-aligned neural-operator representation motivated by this reduced structure. Section~\ref{sec:validation} reports the six-output prediction results, engineering consistency checks, baseline comparisons, and sensitivity studies. Finally, Section~\ref{sec:conclusion} summarizes the findings.

\section{Physical Setup, Data Generation, and Train/Test Split}\label{sec:setup}

As our main benchmark, we consider a planar argon micro-nozzle with an external plume. The DSMC simulations use the following operating and numerical conditions: a fixed inlet pressure of \SI{100}{\kilo\pascal}, a simulation time step \(\Delta t=\num{2e-10}\,\si{\second}\), specular solid walls, and a reference temperature of \SI{273.15}{\kelvin}. The nozzle length is $L=\SI{2.65e-4}{\meter}$. The computational domain is partitioned into two zones: (i) the internal nozzle resolved by a structured grid of $100\times 60$ cells and (ii) the external plume resolved by $40\times 30$ cells. A particle-per-cell (PPC) value of 100 is used; each cell is further subdivided into four subcells (two per direction) to reduce numerical diffusion and improve collision sampling. The inlet-height-based Knudsen number is $\mathrm{Kn}\approx 5\times10^{-4}$ at the entrance and can rise to $\mathrm{Kn}\approx 5\times10^{-3}$ immediately upstream of the normal shock when the outlet back pressure is low (e.g., $P_{\text{back}}=\SI{15}{\kilo\pascal}$). As the back pressure decreases, the shock moves downstream toward the exit and strengthens, which produces steeper streamwise gradients and a wider range of local rarefaction levels.

The dataset used for learning and evaluation is produced on the above two-zone
mesh and stored per case in Tecplot ASCII (\texttt{POINT}) multi-zone format
with consistent \(I\times J\) row ordering. For each outlet back pressure
\(P_{\mathrm{back}}\), we export the standard field columns
\texttt{[X, Y, Density, QX, QY, T, U, V, Txy, Mach, Pressure, Knudsen]}.
The complete back-pressure dataset contains \(N_P=19\) DSMC cases over
\(P_{\mathrm{back}}\in[\SI{15}{\kilo\pascal},\SI{33}{\kilo\pascal}]\),
sampled at \(\SI{1}{\kilo\pascal}\) increments. In the operator-learning
formulation, the outlet back pressure \(P_{\mathrm{back}}\) is the
\emph{branch} input, while the spatial coordinates \((x,y)\), augmented with
shock-aligned features, form the \emph{trunk} input evaluated pointwise on the
grid.

To assess generalization across operating conditions, we adopt a held-out
test protocol at the back-pressure level. Three representative back pressures
are reserved exclusively for testing:
\(P_{\mathrm{back}}=\{16,25,30\}~\mathrm{kPa}\), corresponding to low,
intermediate, and high back-pressure conditions. All remaining pressure cases
are used for training and grouped validation. Validation splits are performed
at the pressure-case level, so that points from the same DSMC field do not
appear simultaneously in training and validation. This design ensures that the
network is evaluated on entirely unseen operating points within the same nozzle
family. The split is physically motivated: decreasing \(P_{\mathrm{back}}\)
drives the internal compression layer downstream and strengthens it, creating
the high-gradient configurations that most strongly test the shock-aligned
inductive bias.

In addition to the back-pressure interpolation test, we also perform a separate
geometry-interpolation test based on throat location. In that study,
\(X_{\mathrm{throat}}/L=0.30\) is held out exclusively for testing. The training
throat ratios are \(0.10\), \(0.15\), \(0.20\), \(0.25\), \(0.35\), \(0.45\),
and \(0.55\).

\subsection{Problem Statement}\label{sec:problem}

Let $\Omega\subset\mathbb{R}^2$ denote the two-dimensional micro-nozzle domain and let 
$\mu=P_{\mathrm{back}}$ be the scalar operating parameter. The objective is to learn a parametric surrogate operator
\begin{equation}
\mathcal{G}_\theta:\; \mu \mapsto \mathbf{q}_\mu(x,y),
\qquad 
\mathbf{q}_\mu=(\rho,U,V,T,\mathrm{Mach},P),
\qquad (x,y)\in\Omega ,
\end{equation}
using DSMC-generated reference fields. Equivalently, the pointwise neural representation is written as
\begin{equation}
f_\theta(P_{\mathrm{back}},x,y)
=
(\rho,U,V,T,\mathrm{Mach},P).
\end{equation}
The surrogate is evaluated in two complementary ways: globally, through relative field errors over the nozzle domain, and locally, through shock-region diagnostics based on density-gradient and velocity-gradient indicators. This distinction is important because small spatial misalignment of the shock can produce large pointwise errors even when the global flow topology is correctly reproduced.

\section{Shock-Centered Low-Rank Structure of the DSMC Compression Layer}
\label{sec:reduced_structure}

\subsection{Density-gradient and gradient-length Knudsen-number diagnostics}
\label{sec:gll_diagnostics}
\begin{figure}[H]
  \centering
  \includegraphics[width=0.75\linewidth]{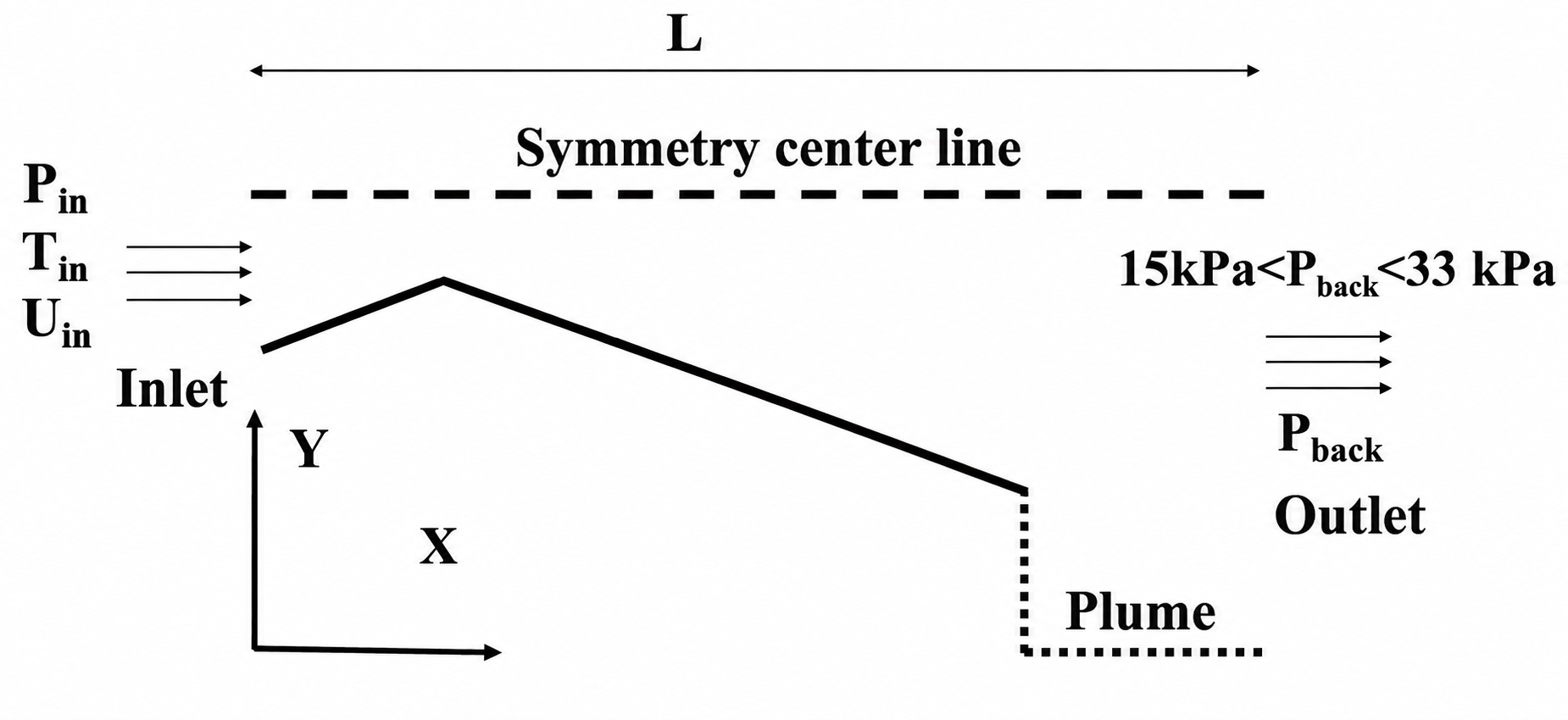}
  \caption{Schematic of the pressure-driven micro-nozzle with downstream plume region.}
  \label{fig:nozzle-schematic}
\end{figure}

As the main DSMC configuration, we consider a pressure-driven planar micro-nozzle with an attached downstream plume region, shown schematically in Fig.~\ref{fig:nozzle-schematic}. 

\FloatBarrier

The inlet total conditions $(P_{\mathrm{in}},T_{\mathrm{in}},U_{\mathrm{in}})$ are prescribed, while the outlet is controlled by the imposed back pressure $P_{\mathrm{back}}$. Specular gas--wall reflections are used, and only the half-domain is simulated by imposing symmetry along the centerline. Because of the small characteristic length and strong area variation, the flow exhibits spatially varying rarefaction and may contain an internal compression layer whose position changes with $P_{\mathrm{back}}$.

The specular wall condition is used as a controlled modeling choice to isolate
the internal compression-layer dynamics. In our earlier DSMC study of
micro/nanoscale converging--diverging nozzles, we showed that gas--surface
interaction strongly affects the shock topology: diffuse reflection enhances
wall-induced viscous and thermal boundary-layer effects, can promote separation,
and may shift the shock system toward the nozzle core, producing
Mach-core/diamond-like shock structures rather than a nearly normal internal
compression layer~\cite{darbandi2011study}. Since the purpose of the present
work is to analyze shock-centered registration and low-rank structure for an
approximately normal internal compression layer, specular reflection is adopted
to reduce wall-accommodation-induced distortion of the core shock. This
idealized boundary condition allows the finite-thickness shock-alignment
mechanism to be examined separately from diffuse-wall boundary-layer effects.

\begin{landscape}
\begin{figure}[H]
    \centering
    \includegraphics[width=0.95\linewidth]{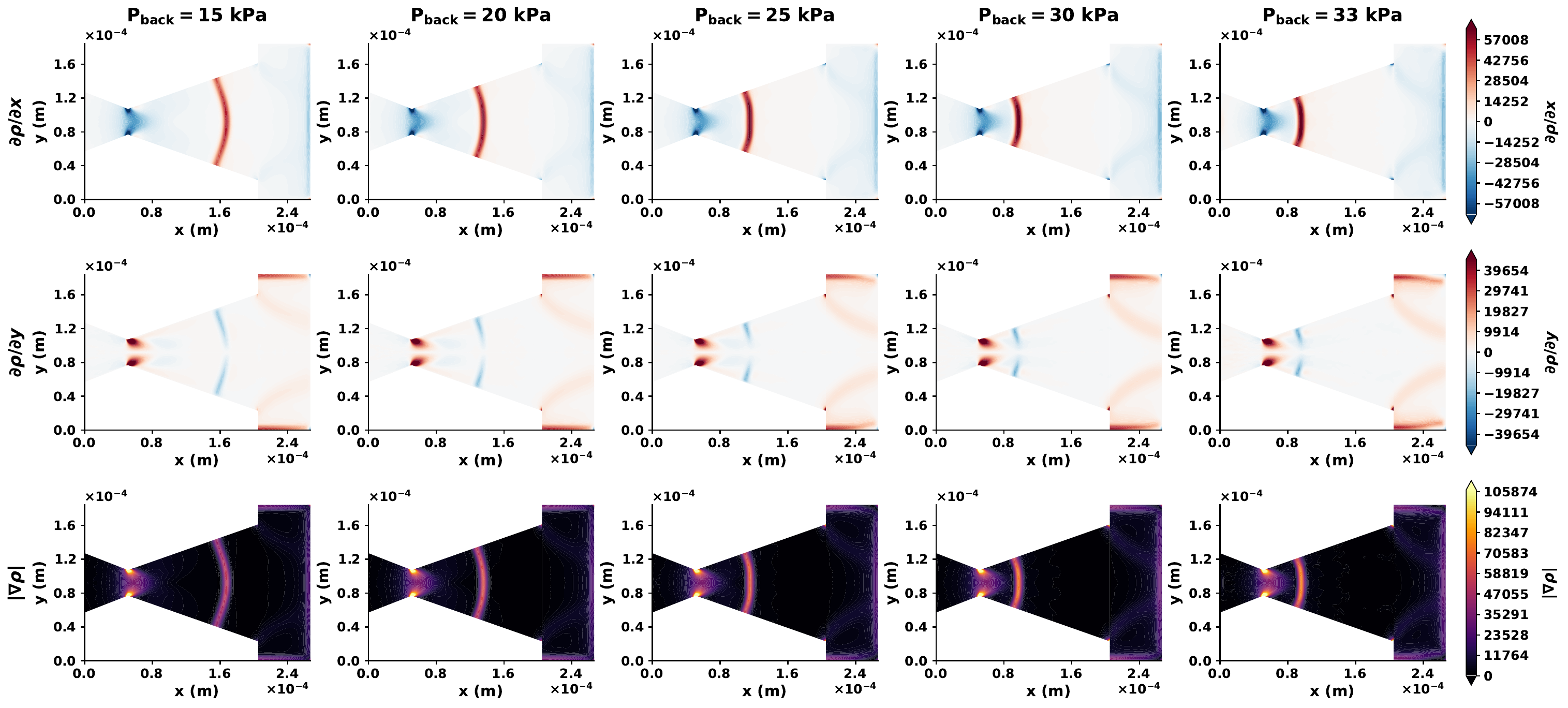}
    \caption{
Density-gradient-based shock indicators computed from the DSMC density field for representative back pressures. 
}
    \label{fig:density_gradient_shock_indicator}
\end{figure}
\end{landscape}

To make the shock topology explicit before evaluating the neural-operator predictions, 
we first examine density-gradient-based shock indicators extracted directly from the DSMC reference fields. 
Although velocity, Mach-number, and pressure contours show the global flow organization, the density-gradient field provides a sharper diagnostic of localized compression layers. 
Accordingly, we compute the streamwise and transverse density-gradient components, 
$\partial \rho/\partial x$ and $\partial \rho/\partial y$, together with the magnitude
\[
|\nabla \rho|=\sqrt{\left(\frac{\partial \rho}{\partial x}\right)^2+
\left(\frac{\partial \rho}{\partial y}\right)^2}.
\]
This diagnostic is particularly useful for rarefied micro-nozzle flows because the shock is not a mathematically discontinuous surface but a finite-thickness rarefied compression layer resolved by the DSMC particle solution.
Figure~\ref{fig:density_gradient_shock_indicator} shows that the high-gradient density band provides a clear and physically interpretable marker of the internal shock position and its displacement with back pressure.

The density-gradient maps reveal several important features of the DSMC reference solution. 
First, the dominant shock indicator is the streamwise density-gradient component, $\partial \rho/\partial x$, which forms a narrow curved band inside the diverging section. 
Second, the transverse component, $\partial \rho/\partial y$, becomes important near the throat corners and along the compression layer, reflecting the two-dimensional adjustment of the flow around the internal shock. 
Third, the magnitude $|\nabla \rho|$ provides the cleanest scalar visualization of the shock footprint and clearly shows the systematic upstream displacement of the compression layer as $P_{\mathrm{back}}$ increases. 
This figure therefore motivates the use of shock-aligned input features in the neural-operator surrogate: the dominant high-gradient structure is localized, parameter-dependent, and moves coherently with the operating back pressure.

To further characterize the local rarefaction and short-gradient-length structure of
the DSMC reference fields, we examine both the conventional Knudsen number and
a gradient-length local Knudsen number. The conventional Knudsen number is
defined as
\[
Kn=\frac{\lambda}{L_{\mathrm{ref}}},
\]
where \(\lambda\) is the local mean free path and \(L_{\mathrm{ref}}\) is the
reference length used in the DSMC simulations, taken here as the inlet height. While this quantity provides
a useful global or reference-scale measure of rarefaction, it does not directly
account for the local length scale of rapid spatial variations. Gradient-length
and continuum-breakdown indicators have therefore been widely used to identify
regions where macroscopic fields vary over kinetic length scales
\cite{swaminathan2016gce_breakdown}.

Therefore, we also compute a gradient-length local (GLL) Knudsen number, denoted by \(Kn_{\mathrm{GLL}}\),
\[
Kn_{\mathrm{GLL}}
=
\max_{\psi\in\{\rho,T,|\mathbf{u}|\}}
\left(
\frac{\lambda}{L_{\psi}}
\right),
\qquad
L_{\psi}
=
\frac{|\psi|+\epsilon_{\psi}}{|\nabla \psi|+\epsilon_{\nabla,\psi}} .
\]
Equivalently, in the numerical implementation we evaluate
\[
Kn_{\mathrm{GLL}}
=
\max_{\psi\in\{\rho,T,|\mathbf{u}|\}}
\left[
\lambda
\frac{|\nabla \psi|+\epsilon_{\nabla,\psi}}
{|\psi|+\epsilon_{\psi}}
\right],
\]
where \(\psi\) denotes density, temperature, or velocity magnitude,
\(|\mathbf{u}|=(U^2+V^2)^{1/2}\). The small regularization constants
\(\epsilon_{\nabla}\) and \(\epsilon_{\psi}\) are used only to avoid division
by nearly zero values and do not affect the resolved high-gradient regions.

\begin{figure*}[t]
  \centering
  \includegraphics[width=0.95\textwidth]{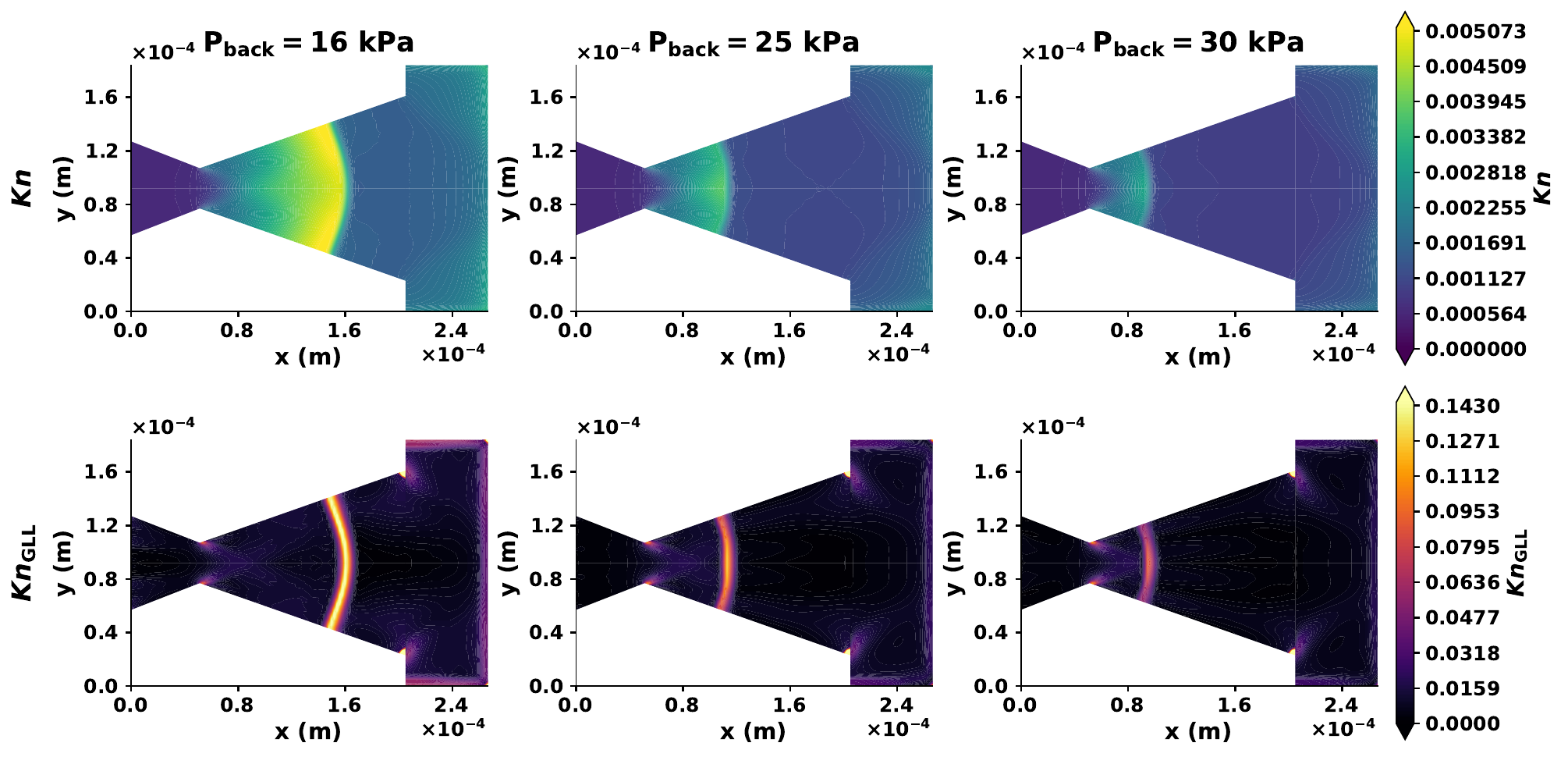}
  \caption{
  Conventional and gradient-length local Knudsen-number contours extracted
  from the DSMC reference fields for the held-out back-pressure cases
  \(P_{\mathrm{back}}=16\), 25, and 30 kPa.}
  \label{fig:knudsen_gll_contours}
\end{figure*}

Figure~\ref{fig:knudsen_gll_contours} compares the conventional Knudsen number
and \(Kn_{\mathrm{GLL}}\) for the held-out back-pressure cases. The conventional
Knudsen number remains small over most of the nozzle--plume domain, with values
on the order of \(10^{-3}\)--\(10^{-2}\). However, \(Kn_{\mathrm{GLL}}\) becomes
substantially larger in localized regions associated with the internal
compression layer, rapid expansion, and downstream plume adjustment. This contrast indicates that the flow is not characterized solely by a large
global Knudsen number; rather, short-gradient-length rarefaction indicators are
concentrated in regions where the macroscopic fields vary rapidly. This
provides additional justification for using DSMC reference data rather than continuum-based solvers. The compression layer identified by \(|\nabla\rho|\) is therefore also the region where the gradient-length Knudsen number and the expected surrogate-model error are most localized. This motivates the following shock-centered analysis, which tests whether the same localized rarefied structure admits a compact reduced representation.

\subsection{Shock-centered reduced representation of the compression layer}
\label{sec:shock_centered_rom}

The density-gradient and gradient-length Knudsen-number diagnostics show that the dominant high-gradient structure is a localized, finite-thickness compression layer. We therefore examine whether this moving structure admits a reduced representation before introducing the neural-operator predictions. 

For each back-pressure case, the centerline compression-layer station is defined from the DSMC density field as
\begin{equation}
x_s(P_{\mathrm{back}})
=
\arg\max_x \left|\frac{\partial \rho}{\partial x}\right|.
\label{eq:xs_density_gradient}
\end{equation}
This density-gradient definition of \(x_s\) is used only for the a posteriori reduced-order analysis of the DSMC database. It is not used to provide oracle shock locations to the neural operator for held-out predictions. During inference, the neural model uses the training-calibrated \(x_s(P_{\mathrm{back}})\) map described in Section~\ref{sec:xs_est}.
The density-gradient definition is used here because the reduced-order analysis focuses on the finite-thickness density compression layer. In the neural-operator implementation, the shock station is estimated from the streamwise velocity-gradient peak. Both indicators identify the same internal compression region within the grid resolution of the present DSMC fields; the density-gradient definition is therefore used for physical diagnostics, while the velocity-gradient definition is retained for the trained alignment map. The consistency between these two shock-station indicators is quantified in
Appendix~\ref{app:xs_consistency}. Across the representative cases, the mean
difference between the density-gradient and velocity-gradient stations is only
\(1.64~\mu\mathrm{m}\), corresponding to \(0.80\) grid spacings, and the maximum
difference is \(2.05~\mu\mathrm{m}\), corresponding to \(1.00\) grid spacing.
An effective layer thickness is estimated from the density jump and the maximum streamwise density gradient,
\begin{equation}
\delta_j
=
\frac{\Delta \rho}{\max |\partial \rho/\partial x|},
\label{eq:delta_jump}
\end{equation}
where \(\Delta\rho\) is evaluated from local upstream and downstream averages around the compression layer. This definition treats the DSMC shock not as a mathematical discontinuity, but as a finite-width kinetic compression layer.

\begin{landscape}
\begin{figure}[p]
\centering
\includegraphics[width=0.75\linewidth]{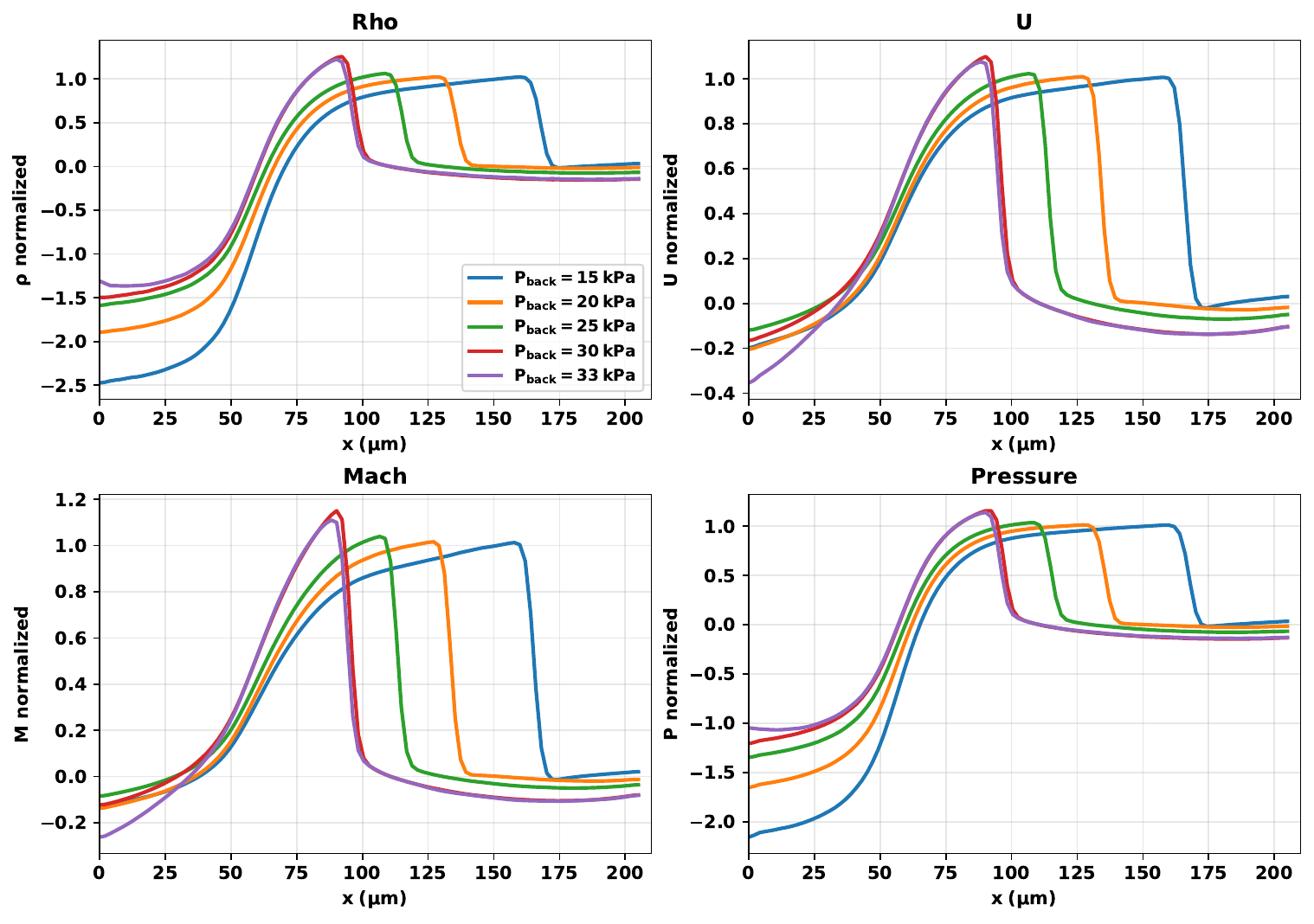}
\caption{
Centerline DSMC profiles in the physical coordinate \(x\) for representative back pressures. 
The dominant effect of varying \(P_{\mathrm{back}}\) is the streamwise displacement of the internal compression layer, which appears as a shift in the sharp transition of density, velocity, Mach number, and pressure. The normalization is performed separately for each back-pressure case to
emphasize the displacement of the compression layer; therefore, the upstream
segments of the normalized profiles are not expected to collapse exactly.
Because the flow is pressure-driven, changing \(P_{\mathrm{back}}\) also
slightly modifies the upstream acceleration and pre-shock state.
}
\label{fig:profiles_x_physical}
\end{figure}
\end{landscape}

\begin{landscape}
\begin{figure}[p]
\centering
\includegraphics[width=0.80\linewidth]{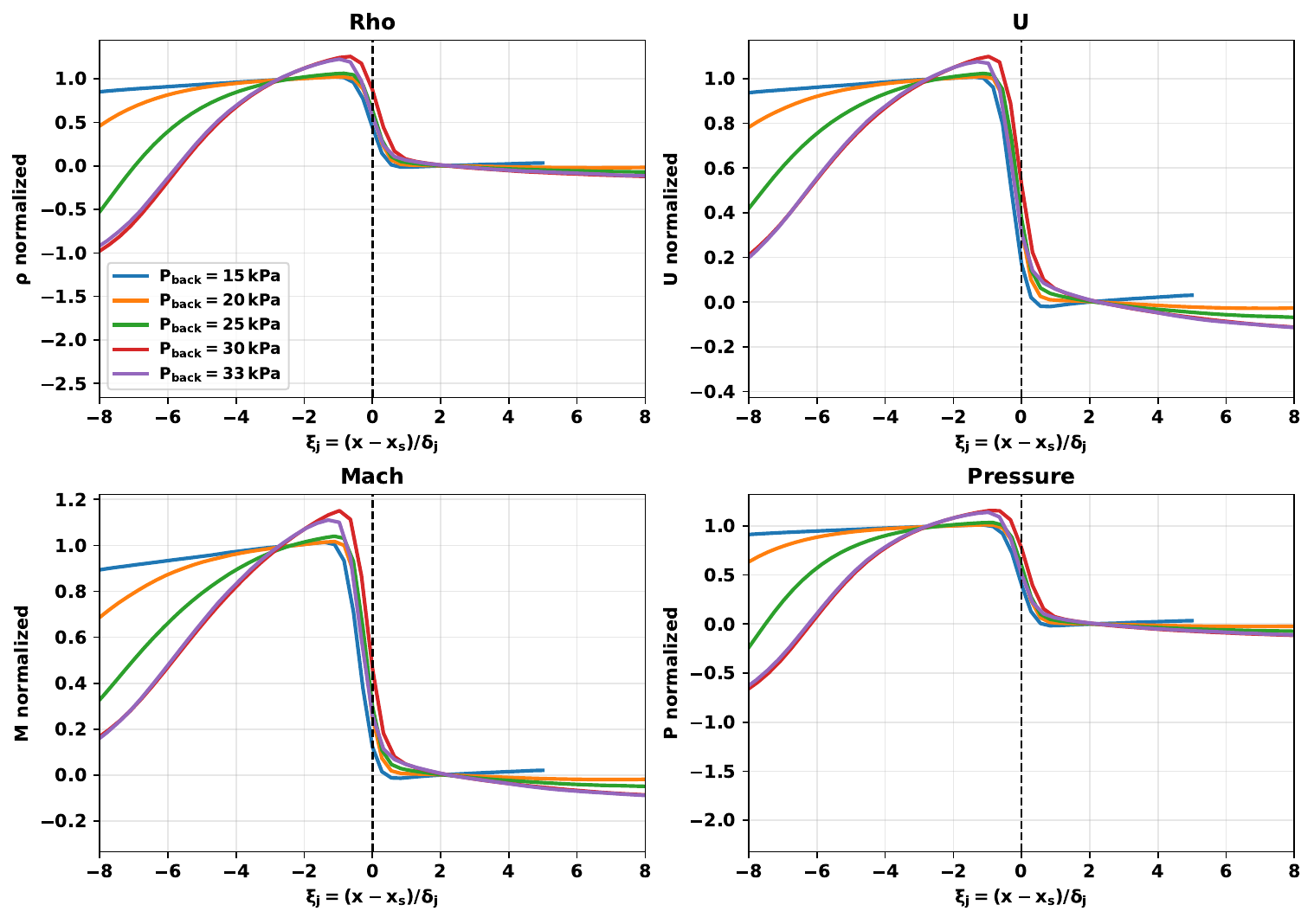}
\caption{
Centerline DSMC profiles expressed in the shock-centered coordinate
\(\xi_j=(x-x_s)/\delta_j\), where \(x_s\) is the compression-layer location and
\(\delta_j=\Delta\rho/\max|\partial\rho/\partial x|\) is the jump-based layer thickness. 
Compared with the physical-coordinate representation, the sharp transition collapses around \(\xi_j=0\), indicating that the dominant parametric variability is largely translational and thickness-scaled.
}
\label{fig:profiles_xi_jump}
\end{figure}
\end{landscape}

Figures~\ref{fig:profiles_x_physical} and \ref{fig:profiles_xi_jump} compare the centerline profiles in the physical coordinate \(x\) and in the shock-centered coordinate
\begin{equation}
\xi_j=\frac{x-x_s}{\delta_j}.
\label{eq:xi_jump}
\end{equation}
In physical space, the profiles are strongly displaced as \(P_{\mathrm{back}}\) changes. After shock-centering and thickness scaling, however, the density, streamwise velocity, Mach number, and pressure profiles collapse in the neighborhood of \(\xi_j=0\). Thus, the dominant back-pressure dependence of the internal compression layer is largely a translation and finite-thickness scaling, rather than a fully high-dimensional deformation of the entire field.

\clearpage

\clearpage

\begin{table}[H]
\centering
\caption{
Compression-layer metrics extracted from the DSMC density-gradient field. The station \(x_s\) is defined by the centerline maximum of \(|\partial\rho/\partial x|\), and \(\delta_j=\Delta\rho/\max|\partial\rho/\partial x|\) is the effective jump-based layer thickness.
}
\label{tab:shock_layer_metrics}
\begin{tabular}{ccccc}
\toprule
\(P_{\mathrm{back}}\) (kPa) 
& \(x_s\) (\(\mu\)m) 
& \(\delta_j\) (\(\mu\)m)
& \(\max|\partial\rho/\partial x|\)
& \(Kn_{\max}\) centerline \\
\midrule
15 & 168.10 & 7.38 & \(5.52\times10^{4}\) & 0.00529 \\
20 & 135.30 & 7.43 & \(6.39\times10^{4}\) & 0.00432 \\
25 & 114.80 & 7.34 & \(6.96\times10^{4}\) & 0.00379 \\
30 & 96.35  & 6.36 & \(7.35\times10^{4}\) & 0.00297 \\
33 & 96.35  & 6.32 & \(7.86\times10^{4}\) & 0.00273 \\
\bottomrule
\end{tabular}
\end{table}
\FloatBarrier

The tabulated shock-layer diagnostics in this subsection are reported for five
representative pressures spanning the back-pressure range. These cases are used
to show the physical trend without overloading the main text; the neural
operator is trained and evaluated using the full pressure dataset described in
Section~\ref{sec:setup}. Table~\ref{tab:shock_layer_metrics} quantifies the same trend. As the back pressure increases from 15 to 33 kPa, the compression layer moves upstream from \(x_s=168.10~\mu\mathrm{m}\) to \(x_s=96.35~\mu\mathrm{m}\), while the effective thickness remains confined to a narrow range, \(\delta_j\approx 6.3\)--\(7.4~\mu\mathrm{m}\). The peak density gradient also increases with back pressure, indicating a sharper compression layer at higher downstream pressure. These trends support the use of a shock-centered coordinate: the primary parametric effect is the displacement and sharpening of a finite-width layer, not a complete reorganization of the centerline flow.

To connect the shock-centered thickness scale to kinetic rarefaction, we compare
\(\delta_j\) with the local mean free path
\(\lambda_s=Kn_s L_{\mathrm{ref}}\) evaluated at the compression-layer station.
The ratio \(\delta_j/\lambda_s\) provides a kinetic measure of the finite shock-layer thickness, while
\(Kn_{\mathrm{GLL},\rho}^{\max}\) identifies whether the same region is also the strongest density-gradient rarefaction zone.
This comparison distinguishes the present analysis from a purely geometric shock registration: the coordinate scale is tied to the DSMC-resolved finite-width kinetic layer.

\begin{table}[H]
\centering
\caption{
Kinetic interpretation of the shock-centered thickness scale.
Here \(\lambda_s=Kn_s L_{\mathrm{ref}}\) is the local mean free path evaluated at the compression-layer station, and \(Kn_{\mathrm{GLL},\rho}^{\max}\) is the maximum density-based gradient-length local Knudsen number along the centerline.
}
\label{tab:prf_kinetic_scale}
\begin{tabular}{ccccccc}
\toprule
\(P_{\mathrm{back}}\) (kPa) 
& \(x_s\) (\(\mu\)m) 
& \(\delta_j\) (\(\mu\)m) 
& \(\lambda_s\) (\(\mu\)m) 
& \(\delta_j/\lambda_s\) 
& \(Kn_s\) 
& \(Kn_{\mathrm{GLL},\rho}^{\max}\) \\
\midrule
15 & 168.10 & 7.38 & 0.134 & 55.27 & \(2.34\times10^{-3}\) & 0.034 \\
20 & 135.30 & 7.43 & 0.133 & 55.74 & \(2.34\times10^{-3}\) & 0.031 \\
25 & 114.80 & 7.35 & 0.121 & 60.95 & \(2.11\times10^{-3}\) & 0.027 \\
30 & 96.35  & 6.36 & 0.113 & 56.28 & \(1.98\times10^{-3}\) & 0.019 \\
33 & 96.35  & 6.32 & 0.082 & 77.16 & \(1.44\times10^{-3}\) & 0.017 \\
\bottomrule
\end{tabular}
\end{table}

Table~\ref{tab:prf_kinetic_scale} shows that the effective compression-layer
thickness is on the order of \(55\)--\(77\) local mean free paths over the
pressure range considered. Thus, the layer is not a discontinuous surface in
the DSMC solution, but a resolved finite-width kinetic structure. The ratio
\(\delta_j/\lambda_s\) should not be interpreted as a universal Boltzmann
shock-thickness constant. Because \(\delta_j\) is extracted from a curved
internal compression layer using centerline density-jump diagnostics, it should
be interpreted as an effective macroscopic registration thickness of the
resolved compression region, not as the molecular thickness of an ideal planar
normal shock. In classical kinetic theory, the thickness of a normal shock in
argon is typically on the order of a few to several tens of local mean free
paths, depending on the upstream Mach number and molecular model. The larger
values obtained here are physically plausible because the present compression
layer is embedded in an accelerating, wall-bounded nozzle flow with area change,
wall confinement, and spatially varying rarefaction. Consequently, the layer
thickness is influenced not only by local collisional relaxation, but also by
the interaction of expansion and compression waves with the confined nozzle
geometry.

This interpretation also explains the observed low-rank collapse. Across the
present back-pressure range, changing \(P_{\mathrm{back}}\) primarily displaces
the compression layer and modifies its effective thickness, while the normalized
internal structure of the layer remains similar when measured in units of
\(\delta_j\). The shock-centered coordinate removes the leading translational
degree of freedom, and the jump-based scaling removes the leading
finite-thickness variation. The remaining variability is associated mainly with
transverse deformation, wall-adjacent gradients, and weak post-layer recovery,
which explains why only a small number of registered POD modes is required.

We also test alternative alignment scales, including the full width at half maximum (FWHM) of the density-gradient peak, a gradient-moment width \(\delta_m\), the density-gradient length \(L_{\rho,s}\), and the local mean free path \(\lambda_s\). These alternatives are used only to assess whether the jump-scaled coordinate provides the most compact representation.

\begin{table}[H]
\centering
\caption{
Compactness of centerline density profiles under progressively aligned coordinates.
\(E_1\) is the energy captured by the leading POD mode, and \(E_{1+2}\) is the cumulative energy captured by the first two modes.
}
\label{tab:coordinate_pod_comparison}
\begin{tabular}{lcc}
\toprule
Coordinate & \(E_1\) & \(E_{1+2}\) \\
\midrule
\(x\) & 83.33\% & 95.25\% \\
\(x-x_s\) & 89.66\% & 98.93\% \\
\((x-x_s)/\delta_j\) & \textbf{98.33\%} & 99.38\% \\
\((x-x_s)/L_{\rho,s}\) & 97.22\% & \textbf{99.79\%} \\
\bottomrule
\end{tabular}
\end{table}
\FloatBarrier

The coordinate comparison in Table~\ref{tab:coordinate_pod_comparison} shows that the apparent dimensionality of the DSMC profiles is strongly coordinate-dependent. In the physical coordinate \(x\), the leading POD mode captures only \(83.33\%\) of the density-profile fluctuation energy. Translating the coordinate by \(x_s\) increases this value to \(89.66\%\), confirming that shock displacement is a major source of apparent variability. The jump-scaled coordinate \(\xi_j=(x-x_s)/\delta_j\) increases the leading-mode energy to \(98.33\%\), while the density-gradient-length scaling \((x-x_s)/L_{\rho,s}\) provides a closely related physics-based alternative. Thus, the DSMC-resolved compression layer becomes nearly low-rank once its motion and finite thickness are accounted for. 
The additional FWHM, gradient-moment, and mean-free-path scalings were also tested, but they did not improve the leading-mode compactness relative to the jump-scaled coordinate. We therefore retain \(\xi_j=(x-x_s)/\delta_j\) as the main reduced coordinate and use \(L_{\rho,s}\) only as a physics-based comparison scale.

\begin{figure}[t]
\centering
\includegraphics[width=0.78\textwidth]{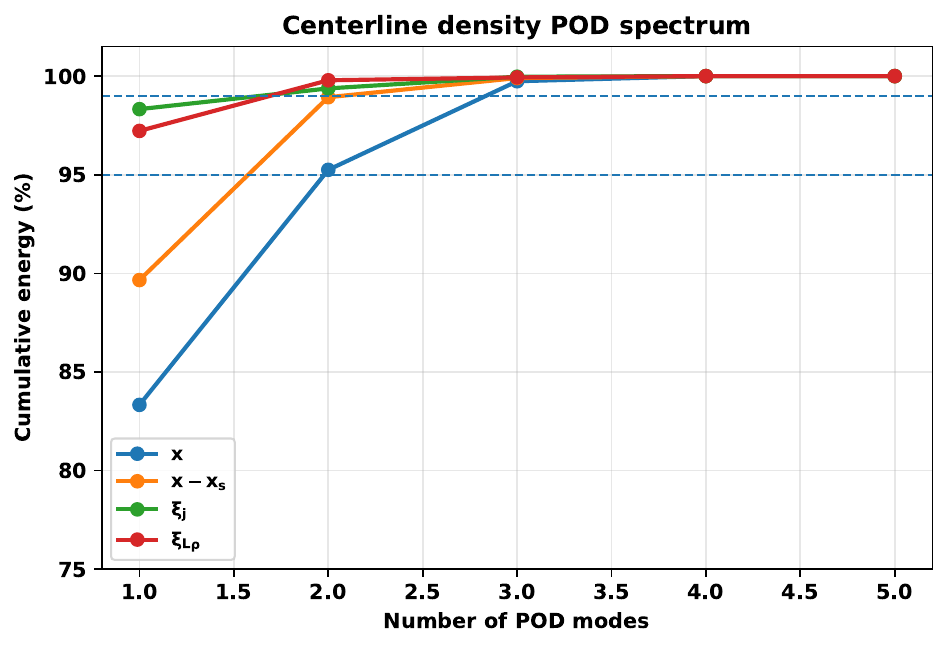}
\caption{
Cumulative POD energy spectrum of the centerline density profiles under
different coordinate representations. The registered coordinates
\(\xi_j=(x-x_s)/\delta_j\) and \((x-x_s)/L_{\rho,s}\) reach high cumulative
energy with fewer modes than the physical coordinate, confirming that
shock-centering and finite-thickness scaling reduce the apparent parametric
dimension of the DSMC compression layer.
}
\label{fig:prf_v7_centerline_pod_spectrum}
\end{figure}

To verify that this conclusion is not a consequence of reporting only the
leading POD mode, we also examine the full cumulative POD spectrum of the
centerline density profiles. Table~\ref{tab:prf_v7_full_pod_spectrum} reports
the number of modes required to capture 95\%, 99\%, and 99.5\% of the
fluctuation energy, while Fig.~\ref{fig:prf_v7_centerline_pod_spectrum}
shows the corresponding cumulative energy curves. In the physical coordinate
\(x\), three modes are required to exceed 99\% energy, whereas the
jump-scaled coordinate \(\xi_j=(x-x_s)/\delta_j\) reaches 95\% with one mode
and 99\% with two modes. The density-gradient-length coordinate
\((x-x_s)/L_{\rho,s}\) behaves similarly. This confirms that the compactness
improvement is not limited to the first POD mode, but reflects a systematic
reduction of the effective dimension after registration.

\begin{table}[t]
\centering
\caption{Full POD compactness of the centerline density profiles. \(N_{95}\), \(N_{99}\), and \(N_{99.5}\) denote the minimum number of modes required to capture 95\%, 99\%, and 99.5\% of the fluctuation energy, respectively.}
\label{tab:prf_v7_full_pod_spectrum}
\begin{tabular}{lccccc}
\toprule
Coordinate & \(E_1\) & \(E_{1+2}\) & \(N_{95}\) & \(N_{99}\) & \(N_{99.5}\) \\
\midrule
\(x\) & 83.33\% & 95.25\% & 2 & 3 & 3 \\
\(x-x_s\) & 89.66\% & 98.93\% & 2 & 3 & 3 \\
\((x-x_s)/\delta_j\) & 98.33\% & 99.38\% & 1 & 2 & 3 \\
\((x-x_s)/L_{\rho,s}\) & 97.22\% & 99.79\% & 1 & 2 & 2 \\
\bottomrule
\end{tabular}
\end{table}
\FloatBarrier

Because the preceding collapse and POD diagnostics are based on centerline
profiles, we further test whether the reduced structure persists in a
two-dimensional shock-window representation. For each DSMC case, the density
field is mapped to a registered coordinate system \((\xi_j,\eta)\), where
\(\xi_j=(x-x_s)/\delta_j\) is the shock-centered streamwise coordinate and
\(\eta\) is the normalized transverse coordinate. This construction removes
the dominant streamwise displacement of the compression layer while retaining
its transverse deformation and wall-adjacent variations.

\begin{table}[t]
\centering
\caption{
Two-dimensional shock-window POD of the density field in the registered
coordinates \((\xi_j,\eta)\). \(N_{95}\), \(N_{99}\), and \(N_{99.5}\) denote
the minimum number of modes required to capture 95\%, 99\%, and 99.5\% of the
fluctuation energy.
}
\label{tab:prf_v7_2d_pod_summary}
\begin{tabular}{ccccc}
\toprule
\(E_1\) & \(E_{1+2}\) & \(N_{95}\) & \(N_{99}\) & \(N_{99.5}\) \\
\midrule
94.98\% & 99.05\% & 2 & 2 & 3 \\
\bottomrule
\end{tabular}
\end{table}
\FloatBarrier

\begin{figure}[t]
\centering
\includegraphics[width=0.78\textwidth]{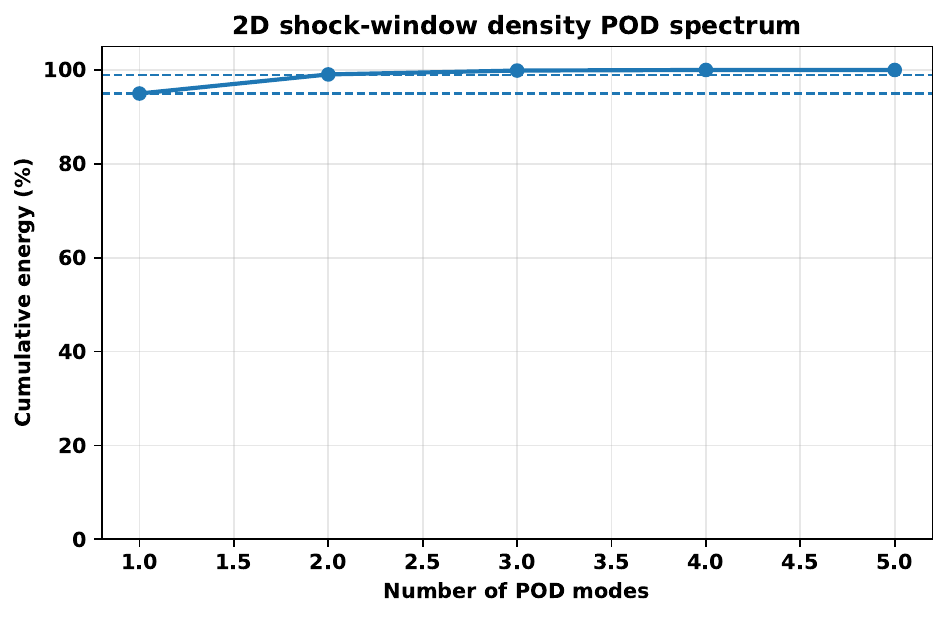}
\caption{
Cumulative POD energy spectrum of the two-dimensional density field in the
registered shock-window coordinate \((\xi_j,\eta)\). The first mode captures
most of the registered shock-window variation, and two modes exceed 99\% of
the fluctuation energy.
}
\label{fig:prf_v7_2d_pod_energy}
\end{figure}

\begin{figure*}[t]
\centering
\includegraphics[width=0.95\textwidth]{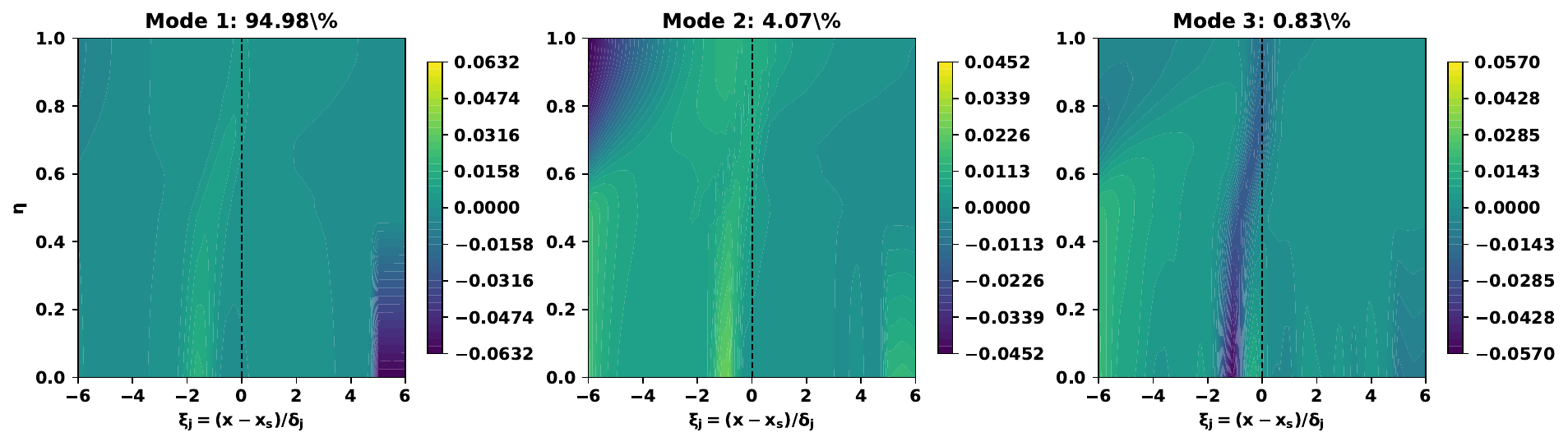}
\caption{
First three POD modes of the two-dimensional density field in the registered
shock-window coordinate \((\xi_j,\eta)\). The leading mode contains the dominant
registered compression-layer structure, while the higher modes are localized
near the finite-thickness layer and wall-adjacent/transverse adjustment
regions. This demonstrates that the shock-centered compactness is not purely a
centerline artifact.
}
\label{fig:prf_v7_2d_pod_modes}
\end{figure*}

The two-dimensional POD results in
Fig.~\ref{fig:prf_v7_2d_pod_energy} and
Fig.~\ref{fig:prf_v7_2d_pod_modes} support the same physical interpretation as
the centerline analysis. The registered shock-window density field remains
compact, with the first mode capturing \(94.98\%\) and the first two modes
capturing \(99.05\%\) of the fluctuation energy. The full-field compactness is
slightly weaker than the centerline compactness, as expected, because the
two-dimensional window retains transverse deformation and corner-induced adjustments. Nevertheless, the rapid decay of the registered
POD spectrum shows that the dominant variability of the DSMC-resolved
compression layer is still governed primarily by shock displacement and
finite-thickness scaling.

\begin{figure*}[t]
\centering
\includegraphics[width=0.78\textwidth]{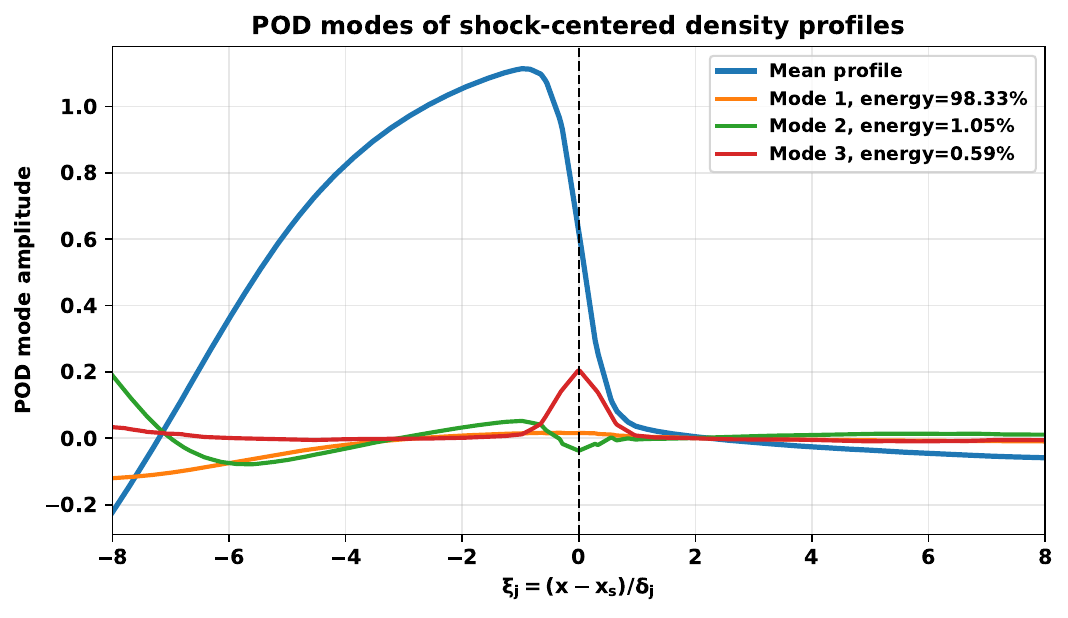}
\caption{
POD modes of the shock-centered density profiles using the jump-scaled coordinate \(\xi_j=(x-x_s)/\delta_j\). The leading mode captures \(98.33\%\) of the fluctuation energy, and the remaining modes are localized mainly near \(\xi_j=0\), showing that the residual variability is concentrated in the finite-thickness compression layer.
}
\label{fig:pod_modes_xi_jump}
\end{figure*}

The POD modes shown in Fig.~\ref{fig:pod_modes_xi_jump} further indicate that
the residual variability after shock-centering is localized mainly around the
finite-thickness compression layer. The purpose of this analysis is not to
restate the well-known displacement of an internal nozzle shock with back
pressure. Rather, it shows that the DSMC-resolved rarefied compression layer has a nearly low-rank structure after registration by a finite-thickness layer scale. This provides a structural explanation for why shock-aligned coordinates are effective for neural-operator prediction.
In this sense, the present analysis can be viewed as a rarefied-flow analogue of moving-shock registration ideas used in reduced-order modeling, but applied here to a DSMC-resolved kinetic compression layer and coupled directly to a shock-aligned neural-operator representation.

This reduced-order analysis provides the physical motivation for the shock-aligned trunk representation used in the neural operator. A model trained only on Cartesian coordinates \((x,y)\) must learn both the smooth pre-/post-shock fields and the pressure-dependent motion of a sharp localized compression layer. This creates a registration problem: a small error in the predicted compression-layer station produces a large pointwise error in \(\rho\), \(U\), Mach number, and \(P\), even if the surrounding flow topology is correct. The signed distance \(d=x-x_s(P_{\mathrm{back}})\), the smooth pre-/post-shock indicator \(s(d)\), and the Gaussian envelopes \(\phi_\sigma(d)\) are therefore not arbitrary empirical features. They are a neural representation of the shock-centered coordinates revealed by the DSMC collapse and POD analysis. In this sense, the trunk representation converts the learning task from reconstructing a moving high-gradient structure in Cartesian space to learning residual variations around an aligned, low-rank compression-layer template.

\section{Shock-Aligned Neural-Operator Representation}
\label{sec:shock_features}

The preceding section showed that the DSMC-resolved compression layer becomes nearly low-rank when expressed in the shock-centered coordinate \(\xi_j=(x-x_s)/\delta_j\). We now use this reduced structure to define the trunk features of the neural operator. The goal is not to impose governing-equation residuals, but to encode the dominant translation and finite-thickness scaling of the compression layer as an inductive bias.

Normal shocks introduce large streamwise gradients and an abrupt change in
flow invariants (e.g., stagnation pressure). Purely data-driven models
tend to smooth such discontinuities when trained with pointwise losses that treat all spatial samples uniformly on
$(x,y)$ alone. We inject an \emph{inductive bias} toward a shock-centered representation
by augmenting the trunk input with features that encode:
(i) signed distance to a training-data-derived shock-station estimate,
(ii) a soft left/right indicator, and (iii) multi-scale proximity via
radial-basis envelopes.

\subsection{Shock station and signed distance}
Let \(\xs=\xs(P_{\mathrm{back}})\) denote the shock \(x\)-location estimated
from the training cases using a monotone affine map of the back pressure.
Define the signed streamwise distance
\begin{equation}
  d(x;P_{\mathrm{back}}) \;=\; x - \xs(P_{\mathrm{back}}),
  \label{eq:signed_distance}
\end{equation}
so that $d<0$ and $d>0$ label pre- and post-shock regions, respectively.
Equation~\eqref{eq:signed_distance} provides a coarse alignment of all
zones with respect to a training-data-derived moving reference associated
with the dominant compression layer.

When training across different $P_{\mathrm{back}}$, the same physical structure (the
shock) appears at different $x$ coordinates. Using $d$ instead of $x$
partly \emph{factorizes} the dependence: the branch stream encodes the
condition ($P_{\mathrm{back}}$), while the trunk sees a frame centered on the estimated compression-layer location. This reduces
the burden on the MLP to learn a large translation in $x$.

\subsection{Soft region indicator}

To inform the model whether a point lies upstream or downstream of the
estimated shock location, we use a smooth sigmoid indicator,
\[
  s(x;P_{\mathrm{back}})
  =
  \frac{1}{1+\exp[-k(x_s(P_{\mathrm{back}})-x)]},
\]
where \(x_s(P_{\mathrm{back}})\) denotes the estimated streamwise location
of the dominant compression layer. With this definition, \(s\approx 1\) upstream of the estimated compression
layer (\(x<x_s\)) and \(s\approx 0\) downstream of it (\(x>x_s\)).

The corresponding pre-/post-shock decomposition can be interpreted
schematically as
\[
  (\rho,U,V,T,\mathrm{Mach},P) \approx 
  F_\mathrm{pre}(x,y)\,s
  \;+\;
  F_\mathrm{post}(x,y)\,\big(1-s\big)
  \;+\;
  \sum_{m} G_m(x,y)\,\phi_{\sigma_m}.
\]
Thus, the pre-shock contribution is dominant upstream of the estimated
shock location, while the post-shock contribution is dominant downstream.
With \(k\) sufficiently large (we use \(k\approx 1.8\times 10^{3}\) in
non-dimensionalized units), the transition layer of \(s\) is thin but
differentiable, which empirically stabilizes optimization compared with
hard binary indicators.

We also tested a hyperbolic-tangent indicator,
\[
  s_{\tanh}
=
\frac{1}{2}\left[1+\tanh\big(\kappa(x_s-x)\big)\right],
\]
which provides a similar smooth transition. In practice, the logistic
form was slightly more robust when combined with feature standardization,
and is therefore used in the final model.

\subsection{Multi-scale shock envelopes}
Shocks are not perfectly discontinuous on a discrete mesh; numerical
dissipation spreads them over a few cells. We capture this spread and
its neighborhood via Gaussian radial-basis envelopes centered at $\xs$:
\begin{equation}
  \phi_\sigma(x;P_{\mathrm{back}}) \;=\;
  \exp\!\left(-\frac{d(x;P_{\mathrm{back}})^2}{2\sigma^2}\right),
  \label{eq:rbf}
\end{equation}
with three scales $\sigma\in\{2\Delta x,\,5\Delta x,\,8\Delta x\}$. Here
$\Delta x$ is a robust spacing estimate (we use the median spacing along each
$y$-row). The small scale focuses on the peak gradient, the intermediate
scale captures the numerically smeared shock thickness, and the large
scale represents the near field where compression/expansion waves interact
with the shock. A dedicated sensitivity study for the Gaussian envelope widths is reported later in Section~\ref{sec:rbf_sensitivity}, where alternative multi-scale tuples are compared under the same held-out back-pressure protocol.

Single-scale envelopes either under-represent the far field (too small
$\sigma$) or blur the shock (too large $\sigma$). A minimal bank of three
scales lets the decoder form constructive/destructive combinations that
resolve both the sharp jump and its footprint, reducing bias without a
large parameter cost.

\subsection{Complete trunk feature vector}
Collecting the primitives described above, the trunk receives
\begin{equation}
t(x,y;P_{\mathrm{back}})
=
[x,y,d,s,|d|,d^2,\phi_{2\Delta x},\phi_{5\Delta x},\phi_{8\Delta x},z_{\mathrm{zone}}].
\label{eq:trunk_vector}
\end{equation}
Including both $d$, $|d|$, and $d^2$ allows the MLP to synthesize
odd/even functions of the signed distance and to approximate narrow
polynomial windows near the interface. \(z_{\mathrm{zone}}\in\{0,1\}\) identifies the internal-nozzle and downstream-plume zones.

\subsection{Scaling and numerical stability}
All trunk channels and the $(\rho,U,V,T,\mathrm{Mach},P)$ targets are standardized with
statistics computed on the \emph{training} split only. This prevents
scale imbalance (e.g., raw $x$ vs.\ unitless $\phi_\sigma$) and
improves conditioning of the LayerNorm activations in the network.
We clamp extremely small $\Delta x$ outliers when estimating the radial-basis-function (RBF) widths to avoid vanishingly thin envelopes:
\[
  \Delta x \;\leftarrow\; \mathrm{median}\!\big(\mathrm{diff}(\mathrm{unique}(x))\big),
  \quad
  \Delta x \ge \Delta x_{\min}.
\]

\subsection{Bias toward correct regularization at shocks}
The feature set in \eqref{eq:trunk_vector} encourages the model to
represent the solution as a \emph{piecewise smooth} function with a
localized transition near $d=0$. Concretely, the decoder learns
expressions of the form
\[
  (\rho,U,V,T,\mathrm{Mach},P) \approx 
  F_\mathrm{pre}(x,y)\,s
  \;+\;
  F_\mathrm{post}(x,y)\,\big(1-s\big)
  \;+\;
  \sum_{m} G_m(x,y)\,\phi_{\sigma_m},
\]
where $F_\mathrm{pre/post}$ and $G_m$ are neural fields. This acts as a shock-localized feature-space regularizer: away from the shock the model reverts to
smooth fields; near $d=0$ it can express sharp yet stable transitions.

\subsection{Estimating the shock station \texorpdfstring{$x_s(P_{\mathrm{back}})$}{xs(Pback)}}
\label{sec:xs_est}

In the present implementation, we use a monotone affine map
\(x_s(P_{\mathrm{back}})=a_0+a_1P_{\mathrm{back}}\), fitted from the
training cases only. For each training pressure, the reference shock station is
identified from the near-centerline location of the maximum streamwise velocity
gradient,
\[
  x_s^{\mathrm{data}}(P_{\mathrm{back}}) \in
  \arg\max_x |\partial U/\partial x|.
\]
For neural-operator prediction, \(x_s(P_{\mathrm{back}})\) is always evaluated from the training-calibrated affine map. For held-out back-pressure cases, the DSMC test field is not used to determine \(x_s\). Therefore, the shock-aligned trunk input does not contain oracle shock-location information from the DSMC solution being predicted.

The role of \(x_s(P_{\mathrm{back}})\) is to provide a training-derived
alignment coordinate, analogous to wall-distance or signed-distance functions
in geometry-aware surrogate models. This coordinate organizes the moving
compression layer into a form that is easier for the trunk network to
represent. The effectiveness of this construction is assessed directly through
the hard-case baseline comparison and the shock-window error metrics reported
below.

\subsection{Fusion--DeepONet Architecture}
\label{sec:architecture}

Classical DeepONet couples branch and trunk representations through a final inner product. In the present formulation, the branch stream ingests only the scalar operating condition \(P_{\mathrm{back}}\), while the trunk stream receives the 10-D shock-aligned feature vector in Eq.~\eqref{eq:trunk_vector}. The two embeddings are combined through elementwise multiplicative fusion followed by a nonlinear decoder. This architecture can be interpreted as a pressure-conditioned MLP with Fusion--DeepONet-style modulation.

\paragraph{Branch stream.}
A small MLP maps the scalar back pressure to a latent vector,
\[
b(P_{\mathrm{back}})\in \mathbb{R}^{d}.
\]

\paragraph{Trunk stream.}
A second MLP maps the shock-aligned trunk feature vector to
\[
g(t)\in \mathbb{R}^{d},
\]
where \(t=t(x,y;P_{\mathrm{back}})\) is defined in Eq.~\eqref{eq:trunk_vector}.

\paragraph{Fusion and decoding.}
The branch and trunk embeddings are fused by a Hadamard product and then decoded into the six macroscopic outputs:
\begin{equation}
z=b(P_{\mathrm{back}})\odot g\!\left(t(x,y;P_{\mathrm{back}})\right),
\qquad
(\rho,U,V,T,\mathrm{Mach},P)=D(z).
\end{equation}
Here, \(D(\cdot)\) is a nonlinear decoder MLP. The multiplicative fusion allows the pressure-conditioned latent representation to modulate the spatial basis functions dimension by dimension, which is useful for representing localized compression layers whose position changes with \(P_{\mathrm{back}}\).
For reproducibility, the exact architectural configuration used in the present study is summarized in Table~\ref{tab:arch_summary}. 

\begin{table*}[t]
\centering
\caption{Common architectural configuration of the proposed Fusion--DeepONet framework. The implemented six-output model uses a fusion width of \(d=128\) and a decoder hidden width of 192.}
\label{tab:arch_summary}
\begin{tabular}{ll}
\toprule
Component / Setting & Configuration \\
\midrule
Branch input & Scalar control parameter: \(P_{\mathrm{back}}\) in the main nozzle study\\
Trunk input (main nozzle case) & 10-D shock-aligned feature vector including $z_{\mathrm{zone}}$ \\
Output variables & $\rho,U,V,T,\mathrm{Mach},P$ \\
Branch network depth & 3 hidden Dense blocks \\
Trunk network depth & 3 hidden Dense blocks \\
Branch hidden width & 128 \\
Trunk hidden width & 128 \\
Fusion width $d$ & 128 \\
Decoder hidden width & 192 \\
Dropout probability & 0.10 \\
Input Gaussian noise & 0.005 \\
Weight regularization & $L_2 = 1\times10^{-5}$ \\
\bottomrule
\end{tabular}
\end{table*}

\subsection{Role of the Shock-Aligned Feature Strategy}

The shock-aligned representation is designed to encode the dominant moving
compression layer observed in the present rarefied micro-nozzle dataset. Its
purpose is to transform the moving high-gradient structure into a more
stationary coordinate description for the trunk network. This is analogous to
using wall-distance or signed-distance coordinates in geometry-aware surrogate
models: information already present in the training fields is reorganized into
a physically meaningful coordinate system. Within the present nozzle family,
the signed-distance coordinate, smooth pre-/post-shock indicator, and
multi-scale Gaussian envelopes provide a compact inductive bias that improves
shock-region reconstruction without imposing governing-equation residuals or
jump conditions. This design directly embeds the reduced-order structure identified in Section~\ref{sec:shock_centered_rom}: the signed distance \(d=x-x_s\) represents the leading translational degree of freedom, while the Gaussian envelopes represent the finite spatial support of the compression layer.
\subsection{Losses and Curriculum Weighting}
The target variables \((\rho,U,V,T,\mathrm{Mach},P)\) are standardized using statistics computed from the training split only. We minimize a weighted Huber loss of the form
\begin{equation}
  \mathcal{L}_{\deltah}(\hat{\mathbf{q}},\mathbf{q})
  =
  \sum_i w_i
  \sum_{m=1}^{6}
  \ell_{\deltah}\!\left(\hat{q}_i^{(m)}-q_i^{(m)}\right),
\end{equation}
where
\[
\mathbf{q}=(\rho,U,V,T,\mathrm{Mach},P).
\]
\begin{equation}
\ell_{\deltah}(r)=
\begin{cases}
  \tfrac12 r^2,& |r|\le \deltah,\\[2pt]
  \deltah\big(|r|-\tfrac12\deltah\big), & |r|>\deltah.
\end{cases}
\end{equation}

Although the proposed features and weighting strategy are motivated by the expected shock topology, the model remains fully data-driven. No governing-equation residuals, conservation constraints, or Rankine--Hugoniot jump conditions are explicitly enforced during optimization. The role of physics in the present framework is therefore to provide an informative inductive bias through feature-space alignment and sample reweighting, rather than to impose hard or soft physical constraints in the loss.

\paragraph{Distance-based weight.} Emphasize shock vicinity with a smooth kernel
\begin{equation}
  W_d(x;P_{\mathrm{back}})=1+\alpha\,\exp\!\Big(-\tfrac{d(x;P_{\mathrm{back}})^2}{2(2\Delta x)^2}\Big),\quad \alpha>0.
\end{equation}

\paragraph{Gradient-based weight.} Estimate $g_i\approx|\partial U/\partial x|$ on each $y$-row
(central difference), normalize by $q_{95}$ (95th percentile), and set
\[
  \tilde g_i=\mathrm{clip}(g_i/q_{95},0,1),\qquad W_g=1+\beta\,\tilde g_i.
\]
The final weight is a convex combination
\begin{equation}
  w_i=\lambda\,W_d(x_i;P_{\mathrm{back}})+(1-\lambda)\,W_g(x_i,y_i),\qquad \lambda\in[0,1].
\end{equation}

\paragraph{Two-phase curriculum.} We train in two stages:
\[
\text{Phase I (Warmup): }\ ({\deltah}_{\text{warm}},\lambda_{\text{warm}}),\quad
\text{Phase II (Focus): }\ ({\deltah}_{\text{focus}},\lambda_{\text{focus}}),
\]
with a smaller Huber threshold in Phase II than in Phase I and stronger gradient emphasis in Phase II.
We use AdamW (weight decay), gradient clipping, ReduceLROnPlateau, and EarlyStopping.
Validation is grouped by $P_{\mathrm{back}}$ (GroupShuffleSplit) to prevent $P_{\mathrm{back}}$ leakage.

\subsection{Training Summary}

For each training pressure, the branch input is the scalar $P_{\mathrm{back}}$, while the trunk input consists of the standardized shock-aligned feature vector in Eq.~\eqref{eq:trunk_vector}. The target vector is $\mathbf{q}=(\rho,U,V,T,\mathrm{Mach},P)$, standardized using statistics from the training split only. Training is performed in two stages: a warmup stage with a larger Huber threshold and a focus stage with stronger emphasis on high-gradient samples. Validation is grouped by $P_{\mathrm{back}}$ to avoid leakage between training and validation operating conditions.

\section{Neural-Operator Validation and Discussion}\label{sec:validation}

\subsection{Six-output field prediction}
\label{sec:six_output_prediction}

Having established that the moving compression layer admits a compact shock-centered representation, we next evaluate the revised surrogate as a six-output operator,
\[
\mathbf{q}=(\rho,U,V,T,\mathrm{Mach},P),
\]
rather than as a velocity-only or four-output model. 
This extension is important because an engineering surrogate for micro-nozzle flow should predict the density, both velocity components, temperature, Mach number, and pressure with comparable fidelity.
 The validation is interpreted here as a predictive test of the reduced
structure identified in Section~\ref{sec:shock_centered_rom}. If the dominant
parametric variability of the DSMC fields is indeed governed by translation
and finite-thickness scaling of the compression layer, then a neural operator
whose trunk features encode \(x-x_s\), smooth pre-/post-layer separation, and
localized shock envelopes should improve the shock-window prediction relative
to non-aligned or weakly aligned baselines. The baseline and sensitivity
studies below are therefore not presented only as architecture comparisons;
they test whether the shock-centered compactness observed in the DSMC database
translates into improved prediction for unseen operating conditions. The held-out cases \(P_{\mathrm{back}}=16\), 25, and 30 kPa are evaluated over both DSMC zones combined. We report global relative \(L_2\) errors as the primary quantitative metric and use contour-error maps only as spatial diagnostics of where the remaining discrepancies are localized. The resulting global relative \(L_2\) errors for all six quantities are summarized in Table~\ref{tab:six_output_global_errors}.

\begin{table}[t]
\centering
\caption{
Global relative \(L_2\) errors of the revised six-output surrogate over both DSMC zones. 
The surrogate predicts \(\mathbf{q}=(\rho,U,V,T,\mathrm{Mach},P)\) for the held-out back-pressure cases.
}
\label{tab:six_output_global_errors}
\begin{tabular}{lccc}
\toprule
Quantity & \(P_{\mathrm{back}}=16\) kPa & \(P_{\mathrm{back}}=25\) kPa & \(P_{\mathrm{back}}=30\) kPa \\
\midrule
\(\rho\)        & 4.8\%  & 3.3\% & 3.3\% \\
\(U\)           & 11.2\% & 7.5\% & 8.9\% \\
\(V\)           & 14.8\% & 7.6\% & 5.3\% \\
\(T\)           & 4.3\%  & 2.0\% & 1.9\% \\
\(\mathrm{Mach}\) & 13.2\% & 8.7\% & 10.0\% \\
\(P\)           & 6.8\%  & 3.8\% & 3.8\% \\
\bottomrule
\end{tabular}
\end{table}

The global errors show that density, temperature, and pressure are predicted with relatively low errors, especially for the intermediate and high back-pressure cases. The largest errors occur for the low-back-pressure case, \(P_{\mathrm{back}}=16\) kPa, particularly in the shock-sensitive kinematic quantities \(U\), \(V\), and Mach number. This behavior is physically expected because the low-pressure case lies close to the lower end of the pressure range and contains a stronger, downstream-shifted compression layer. A small streamwise offset in the predicted shock location therefore produces amplified pointwise differences, even when the global flow topology is well reproduced.

\clearpage
\begin{landscape}
\begin{figure}[p]
\centering
\includegraphics[
    width=0.98\linewidth,
    height=0.82\textheight,
    keepaspectratio
]{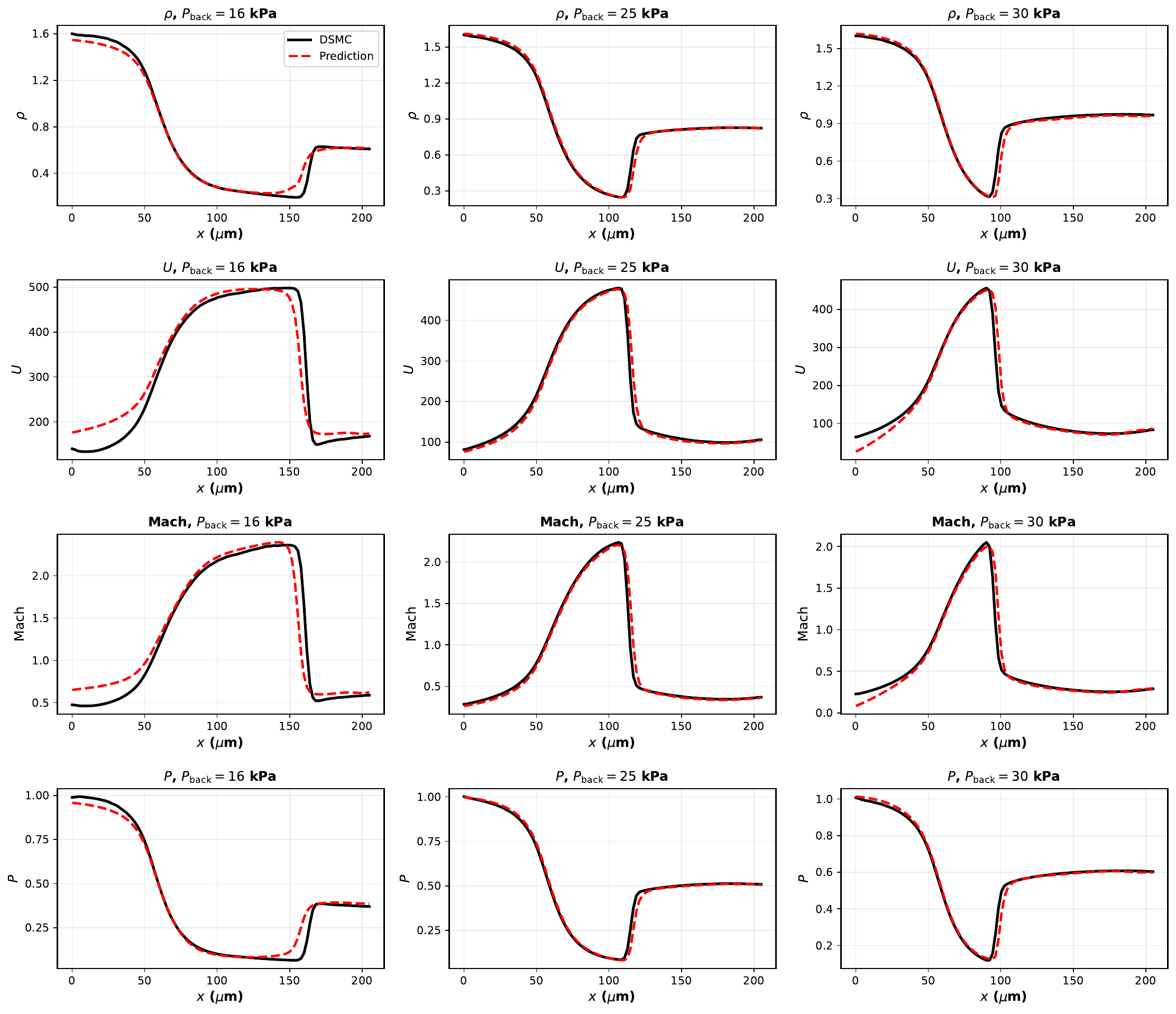}
\caption{
Centerline comparison between DSMC reference fields and the six-output shock-aligned Fusion--DeepONet predictions for the held-out back-pressure cases. The profiles are extracted along the symmetry line of the internal nozzle zone. The model captures the main acceleration, shock-induced drop, and downstream recovery trends; the low-back-pressure case highlights the expected sensitivity of pointwise
velocity and Mach-number errors to small shifts in the compression-layer
location.
}
\label{fig:centerline_profiles_6out}
\end{figure}
\end{landscape}
\clearpage

Figure~\ref{fig:centerline_profiles_6out} provides a stricter one-dimensional diagnostic than global error norms alone. The surrogate reproduces the centerline trends in density, velocity, Mach number, and pressure for all three held-out cases. The agreement is particularly strong for \(P_{\mathrm{back}}=25\) and \(30\) kPa. For \(P_{\mathrm{back}}=16\) kPa, the dominant discrepancy is not a global failure of the prediction but a small offset in the shock station, which shifts the sharp drop in \(U\), Mach number, and pressure. This explains why the local pointwise errors near the shock are larger than the global field errors.

\begin{landscape}
\begin{figure}[p]
\centering
\includegraphics[
width=1.05\linewidth,
trim=0 0 0 1.0cm,
clip
]{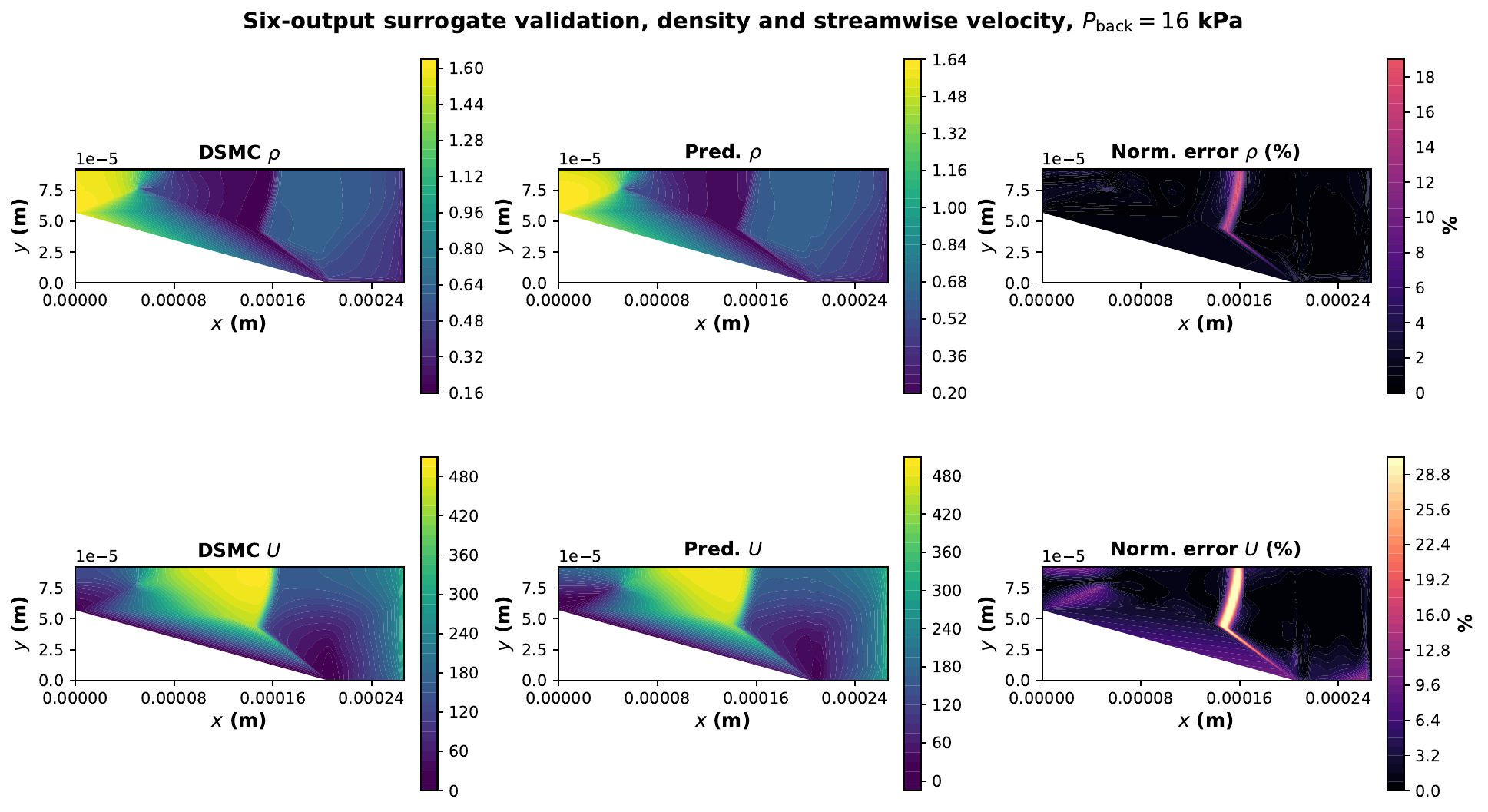}
\caption{
Six-output surrogate validation for the most challenging low-back-pressure held-out case,
\(P_{\mathrm{back}}=16~\mathrm{kPa}\), part I: density and streamwise velocity. Each row compares the DSMC reference field, the surrogate prediction, and the normalized error diagnostic. The error maps show
\(E_q=|\hat q-q|/(q_{99}-q_1)\), clipped for visualization.
}
\label{fig:main_contours_pr16_part1}
\end{figure}
\end{landscape}
\clearpage

\begin{landscape}
\begin{figure}[p]
\centering
\includegraphics[
width=1.05\linewidth,
trim=0 0 0 1.0cm,
clip
]{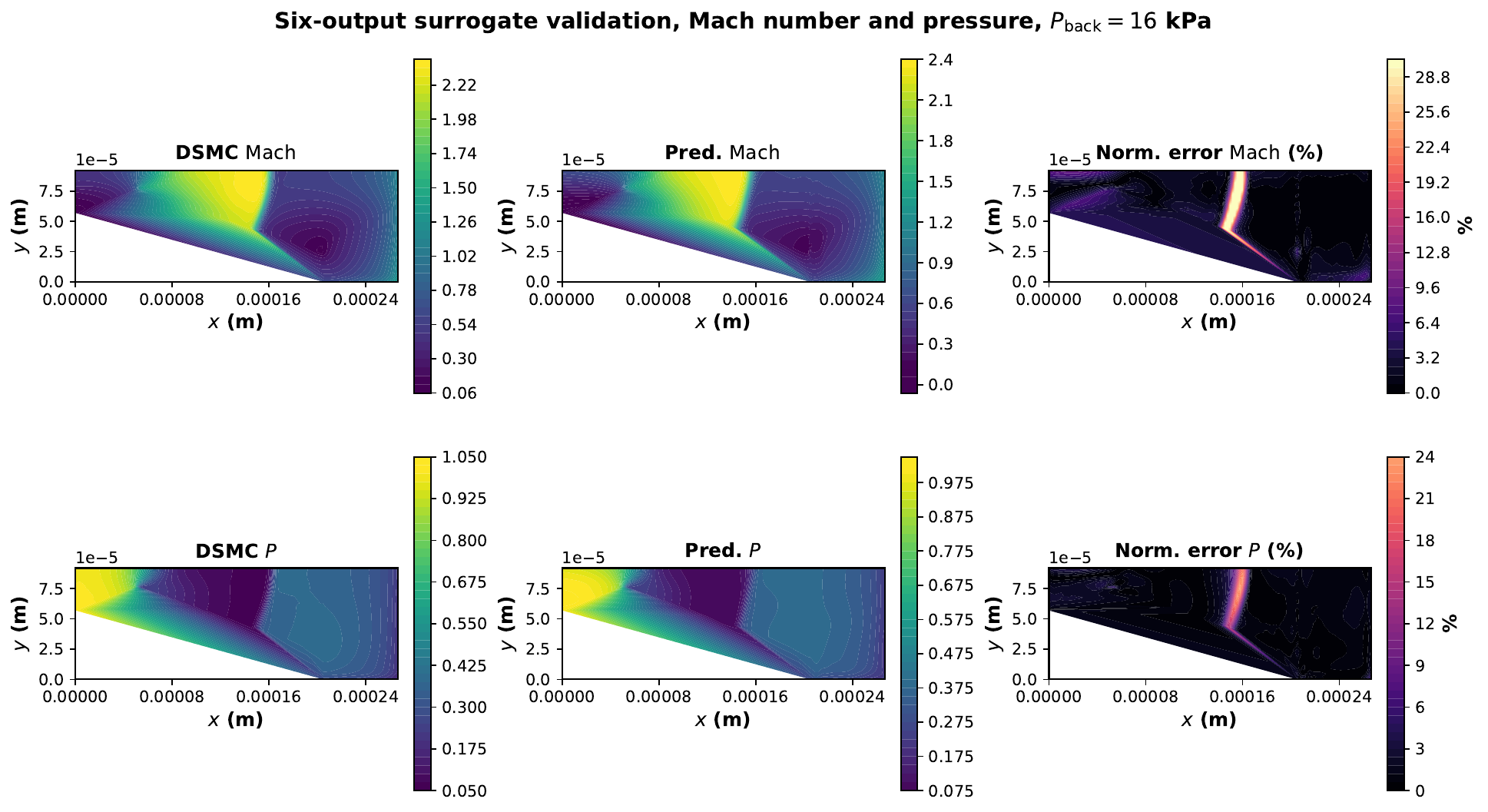}
\caption{
Six-output surrogate validation for \(P_{\mathrm{back}}=16~\mathrm{kPa}\), part II: Mach number and pressure. The largest discrepancies remain localized near the compression layer, indicating that the dominant error source is a small shock-location offset rather than a global failure of the field reconstruction.
}
\label{fig:main_contours_pr16_part2}
\end{figure}
\end{landscape}
\clearpage

\begin{landscape}
\begin{figure}[p]
\centering
\includegraphics[
width=1.05\linewidth,
trim=0 0 0 1.0cm,
clip
]{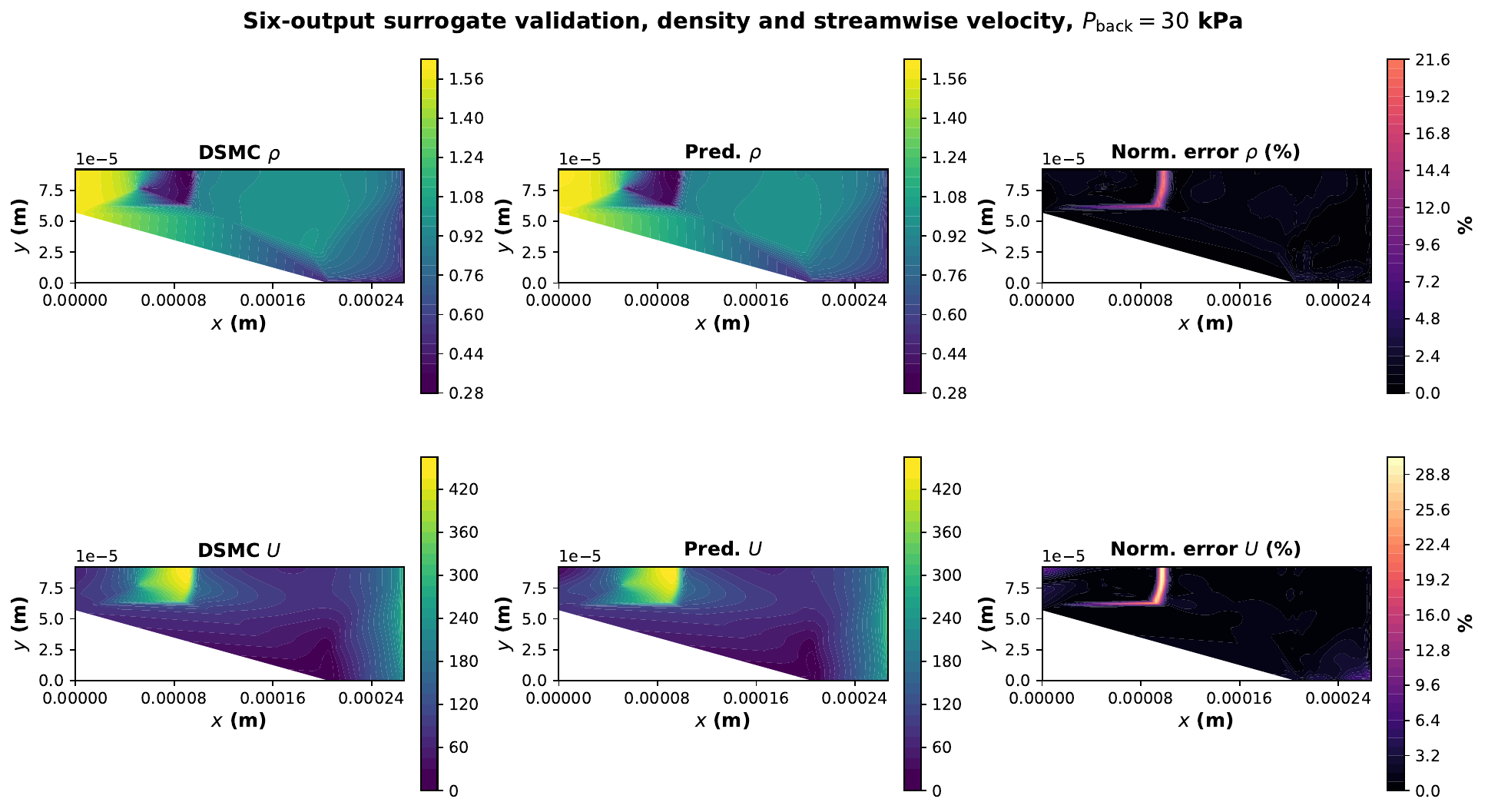}
\caption{
Six-output surrogate validation for the high-back-pressure held-out case,
\(P_{\mathrm{back}}=30~\mathrm{kPa}\), part I: density and streamwise velocity.
}
\label{fig:main_contours_pr30_part1}
\end{figure}
\end{landscape}
\clearpage

\begin{landscape}
\begin{figure}[p]
\centering
\includegraphics[
width=1.05\linewidth,
trim=0 0 0 1.0cm,
clip
]{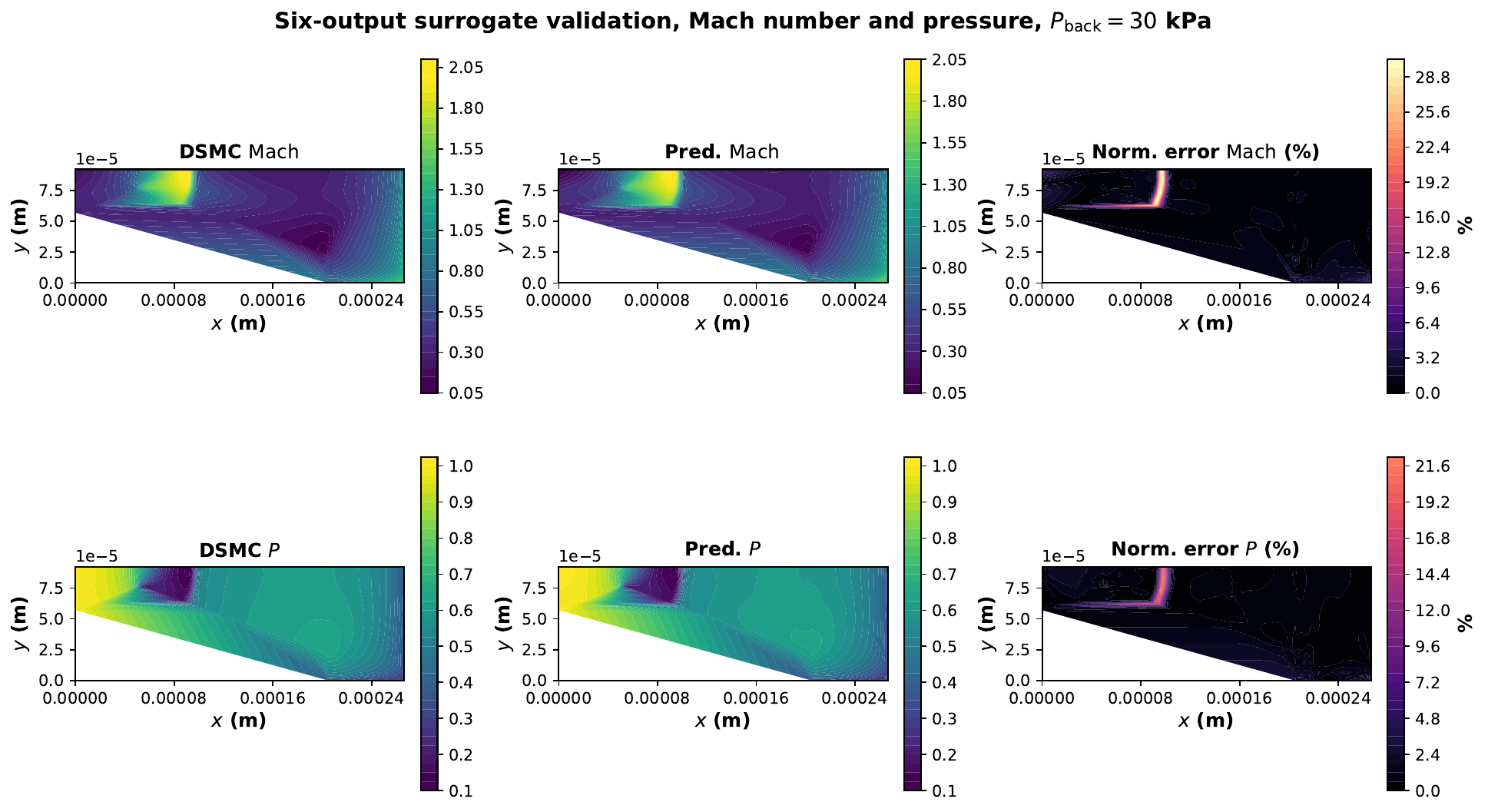}
\caption{
Six-output surrogate validation for \(P_{\mathrm{back}}=30~\mathrm{kPa}\), part II: Mach number and pressure. The DSMC and predicted fields show close agreement in the global flow topology, while the normalized error maps indicate localized discrepancies near high-gradient regions.
}
\label{fig:main_contours_pr30_part2}
\end{figure}
\end{landscape}
\clearpage

The two-dimensional comparisons in Figs.~\ref{fig:main_contours_pr16_part1}--\ref{fig:main_contours_pr16_part2}
and Figs.~\ref{fig:main_contours_pr30_part1}--\ref{fig:main_contours_pr30_part2} confirm that the surrogate captures the main shock topology and post-shock field organization. The residual errors are spatially localized: they appear primarily as thin bands near the compression layer and near outlet-adjacent high-gradient regions. This localization is important because it shows that the largest pointwise discrepancies arise from shock-position sensitivity, while the overall field structure remains well reconstructed.

\subsection{Exit-plane engineering diagnostics}
\label{sec:exit_engineering_metrics}

In addition to field-level shock visualization, we also extract several exit-plane engineering quantities from the DSMC reference solutions. These exit-plane quantities are not the main focus of the present work; they are included only to verify that the shock-centered representation does not degrade integrated nozzle responses. These diagnostics provide a compact measure of how the imposed back pressure modifies the nozzle outflow and help distinguish global performance trends from local contour-level errors. All quantities in this subsection are evaluated at the right boundary of the main nozzle zone, i.e., at the interface between the internal nozzle and the downstream plume/buffer region. It should be emphasized that \(P_{\mathrm{back}}\) denotes the imposed downstream boundary pressure in the plume/buffer region, whereas \(\overline{P}_e\) is the area-averaged static pressure evaluated at the main-nozzle exit plane. These two pressures are therefore not expected to be identical, because the exit plane is upstream of the imposed back-pressure boundary and is connected to it through the downstream plume region. Since the computation is two-dimensional and only a half-domain is simulated using symmetry, the reported values are interpreted as symmetry-corrected quantities per unit depth.

The exit mass flow rate per unit depth is computed as
\begin{equation}
    \dot{m}' = \int_{A_e} \rho U \, dy ,
    \label{eq:exit_mdot}
\end{equation}
where $A_e$ denotes the vertical exit plane of the main nozzle zone and $U$ is the streamwise velocity. The exit-plane momentum and pressure contributions to the thrust-equivalent flux are evaluated as
\begin{equation}
    F'_{\mathrm{mom}} = \int_{A_e} \rho U^2 \, dy ,
    \label{eq:exit_momentum_force}
\end{equation}
and
\begin{equation}
    F'_{\mathrm{press}} = \int_{A_e} \left(P - P_{\mathrm{back}}\right) dy ,
    \label{eq:exit_pressure_force}
\end{equation}
so that the total exit-plane force per unit depth is
\begin{equation}
    F'_{\mathrm{exit}} = F'_{\mathrm{mom}} + F'_{\mathrm{press}} .
    \label{eq:exit_total_force}
\end{equation}
These quantities should be interpreted as exit-plane flux measures rather than the fully integrated mechanical force on the nozzle walls. Their purpose here is to characterize the pressure-conditioned outflow state and to provide additional reference quantities for assessing surrogate-model consistency.

To complement the field-wise error metrics, we evaluate several engineering quantities extracted from the predicted exit-section fields. These quantities include the mass flow rate per unit depth, the mass-flux-weighted exit Mach number, the area-averaged exit pressure, and the momentum, pressure, and total thrust-equivalent exit-plane fluxes per unit depth. Table~\ref{tab:engineering_metrics_nn} compares these quantities between the DSMC reference fields and the six-output surrogate predictions for the held-out back-pressure cases. The results show that the surrogate preserves the integrated engineering response with high accuracy, even when localized shock-position errors are present in the field contours.

\begin{landscape}
\begin{table*}[t]
\centering
\caption{
Exit-section engineering quantities computed from DSMC reference fields and six-output surrogate predictions for the held-out back-pressure cases. The table includes the mass flow rate per unit depth, mass-flux-weighted exit Mach number, area-averaged exit pressure, and thrust-equivalent exit-plane flux components. The relative difference is defined as
\(100(\mathrm{Surrogate}-\mathrm{DSMC})/\mathrm{DSMC}\).
}
\label{tab:engineering_metrics_nn}
\scriptsize
\begin{tabular}{llrrrrrrrrr}
\toprule
& & \multicolumn{3}{c}{\(P_{\mathrm{back}}=16\) kPa} 
& \multicolumn{3}{c}{\(P_{\mathrm{back}}=25\) kPa} 
& \multicolumn{3}{c}{\(P_{\mathrm{back}}=30\) kPa} \\
\cmidrule(lr){3-5}\cmidrule(lr){6-8}\cmidrule(lr){9-11}
Quantity & Unit & DSMC & Surrogate & Diff. & DSMC & Surrogate & Diff. & DSMC & Surrogate & Diff. \\
\midrule
\(\dot{m}'\) & kg\,s\(^{-1}\)\,m\(^{-1}\)
& \(9.120\times10^{-3}\) & \(8.701\times10^{-3}\) & \(-4.60\%\)
& \(8.770\times10^{-3}\) & \(8.794\times10^{-3}\) & \(0.26\%\)
& \(8.777\times10^{-3}\) & \(8.829\times10^{-3}\) & \(0.59\%\) \\

\(\overline{M}_{e,\dot m}\) & --
& 0.482 & 0.461 & \(-4.33\%\)
& 0.346 & 0.342 & \(-1.22\%\)
& 0.299 & 0.294 & \(-1.80\%\) \\

\(\overline{P}_e\) & kPa
& 37.554 & 38.323 & \(2.05\%\)
& 51.730 & 51.690 & \(-0.08\%\)
& 61.148 & 60.597 & \(-0.90\%\) \\

\(F'_{\mathrm{mom}}\) & N\,m\(^{-1}\)
& 1.233 & 1.117 & \(-9.40\%\)
& 0.844 & 0.839 & \(-0.63\%\)
& 0.723 & 0.720 & \(-0.38\%\) \\

\(F'_{\mathrm{press}}\) & N\,m\(^{-1}\)
& 2.974 & 3.081 & \(3.57\%\)
& 3.689 & 3.683 & \(-0.15\%\)
& 4.298 & 4.222 & \(-1.77\%\) \\

\(F'_{\mathrm{exit}}\) & N\,m\(^{-1}\)
& 4.207 & 4.197 & \(-0.23\%\)
& 4.533 & 4.522 & \(-0.24\%\)
& 5.021 & 4.942 & \(-1.57\%\) \\
\bottomrule
\end{tabular}
\end{table*}
\end{landscape}

The integrated quantities in Table~\ref{tab:engineering_metrics_nn} confirm that the surrogate predictions are not only locally consistent with the DSMC fields but also preserve engineering-relevant exit-plane responses. The mass flow rate per unit depth is recovered within \(4.6\%\) for the most challenging low-back-pressure case and within \(1\%\) for the 25 and 30 kPa cases. The total exit-plane force per unit depth is predicted with errors below \(2\%\) for all three held-out cases. The largest individual discrepancy occurs in the momentum contribution for \(P_{\mathrm{back}}=16~\mathrm{kPa}\), which is consistent with the localized shock-position offset observed in the centerline and contour diagnostics. Nevertheless, the total exit-plane force remains accurate because the pressure contribution is well captured and partially compensates the momentum-flux discrepancy.

\subsection{Hard-case comparison with baseline neural operators}
\label{sec:hardcase_model_comparison}

To further isolate the contribution of the proposed shock-aligned Fusion--DeepONet architecture, we perform a hard-case comparison at \(P_{\mathrm{back}}=16~\mathrm{kPa}\), which is the most shock-sensitive held-out condition in the present study. The proposed model is compared against three baselines trained on the same DSMC cases: a vanilla pointwise multilayer perceptron (MLP), a vanilla DeepONet without shock-aligned trunk features, and a zone-wise Fourier neural operator (FNO) baseline. All models are evaluated on the same six predicted quantities,
\[
\mathbf{q}=(\rho,U,V,T,\mathrm{Mach},P).
\]
The comparison includes global field errors, shock-window errors, gradient-weighted errors, and the shock-location error inferred from the streamwise velocity-gradient peak in the main-nozzle zone. The shock-location error reported here is used only as an a posteriori
diagnostic of the predicted field and is not supplied to the model during prediction.

\begin{table*}[t]
\centering
\caption{
Hard-case model comparison for the held-out \(P_{\mathrm{back}}=16~\mathrm{kPa}\) case. 
The table reports global relative \(L_2\) error averaged over the six predicted quantities, shock-window relative \(L_2\) error, gradient-weighted relative \(L_2\) error, and shock-location error estimated from the streamwise velocity gradient in the main-nozzle zone.
}
\label{tab:hardcase_model_comparison}
\small
\resizebox{\textwidth}{!}{%
\begin{tabular}{lccccc}
\toprule
Model 
& Global mean 
& Shock-window mean 
& Grad.-weighted mean 
& \(U\) shock-window 
& \(|\Delta x_s|\) \((\mu\mathrm{m})\) \\
\midrule
Vanilla MLP 
& 4.86\% 
& 9.75\% 
& 7.34\% 
& 8.79\% 
& 2.05 \\

Vanilla DeepONet 
& 6.66\% 
& 22.27\% 
& 8.34\% 
& 21.19\% 
& 2.05 \\

FNO 
& 5.61\% 
& 17.52\% 
& 5.99\% 
& 17.65\% 
& 2.05 \\

Shock-aligned Fusion--DeepONet 
& \textbf{3.15\%} 
& \textbf{4.51\%} 
& \textbf{3.53\%} 
& \textbf{3.45\%} 
& \textbf{0.00} \\
\bottomrule
\end{tabular}%
}
\end{table*}

Table~\ref{tab:hardcase_model_comparison} shows that the shock-aligned Fusion--DeepONet provides the lowest error across all three summary metrics. The improvement is especially pronounced in the shock-window and gradient-weighted metrics, which directly emphasize the moving compression layer. Compared with the vanilla DeepONet and FNO baselines, the proposed model substantially reduces the shock-window error and eliminates the shock-location offset measured from the velocity-gradient peak. This confirms that the main advantage of the proposed model does not arise solely from network capacity, but from aligning the trunk representation with the dominant shock topology.

\begin{figure*}[t]
\centering
\includegraphics[width=0.92\textwidth]{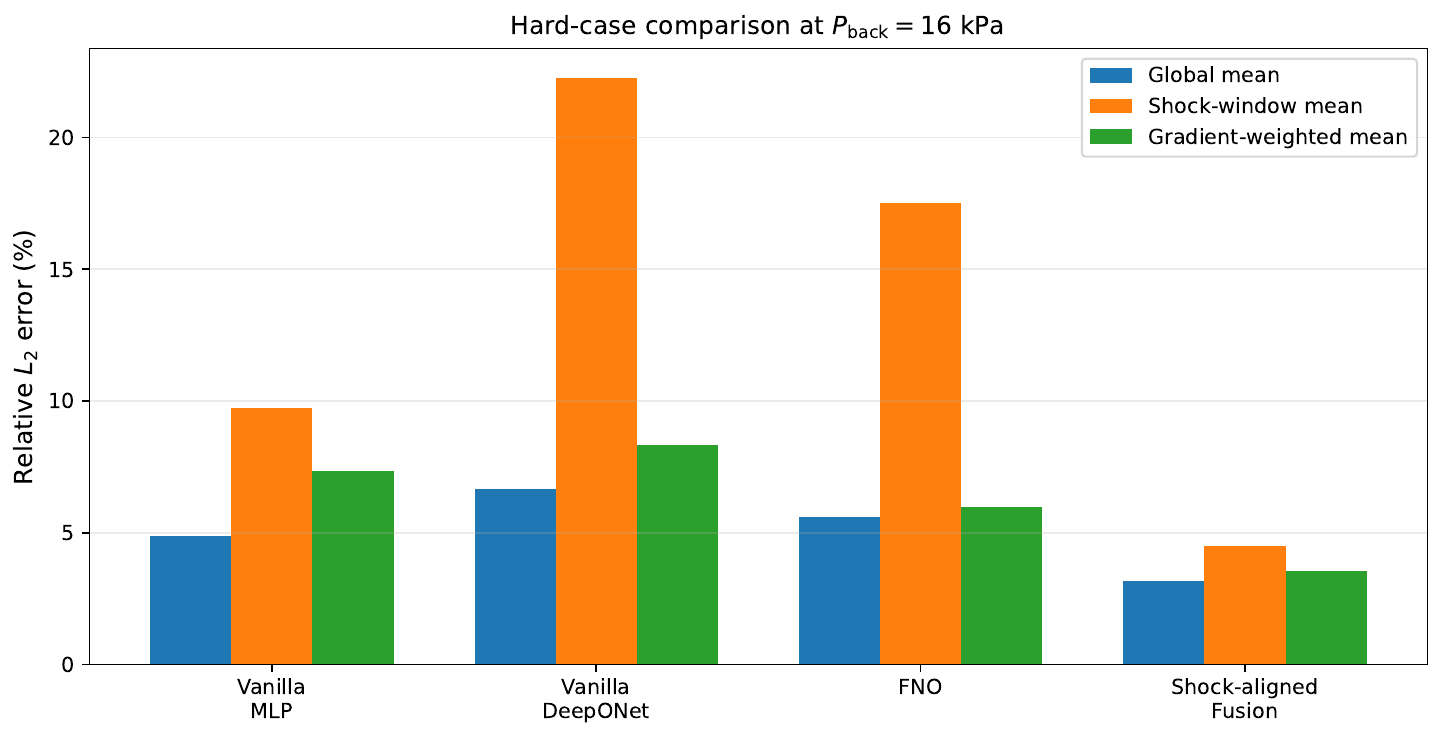}
\caption{
Hard-case comparison of baseline models and the proposed shock-aligned Fusion--DeepONet
for the held-out \(P_{\mathrm{back}}=16~\mathrm{kPa}\) case. The bars report the global mean relative \(L_2\) error, the shock-window mean relative \(L_2\) error, and the gradient-weighted mean relative \(L_2\) error over the six predicted quantities. The largest improvement occurs in the shock-localized metrics, showing that the main benefit of the shock-aligned representation is concentrated near the moving finite-thickness compression layer rather than uniformly over the full field.
}
\label{fig:hardcase_summary_bars}
\end{figure*}

Figure~\ref{fig:hardcase_summary_bars} visualizes the same trend: the global errors of all models are moderate, but the difference becomes much clearer when the metric focuses on the shock region. This supports the central design choice of the proposed approach, namely that a pressure-conditioned neural operator benefits from explicit shock-aligned trunk features when the dominant error source is the displacement of a localized compression layer.

\begin{figure*}[t]
\centering
\includegraphics[width=0.95\textwidth]{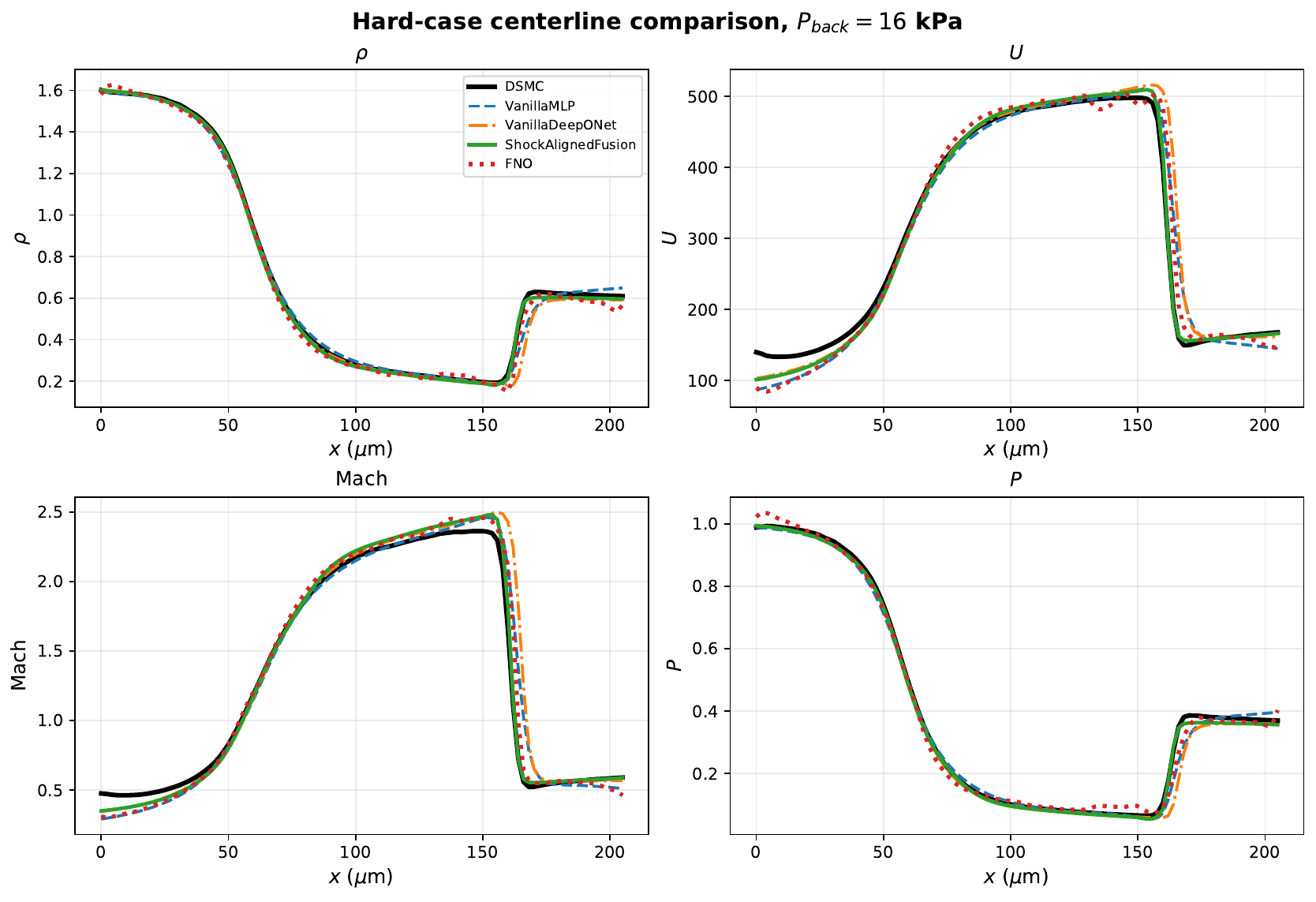}
\caption{
Centerline comparison of DSMC data, baseline neural surrogates, and the proposed shock-aligned Fusion--DeepONet for the hard held-out case \(P_{\mathrm{back}}=16~\mathrm{kPa}\). The proposed model better aligns the shock-induced drop in \(U\), Mach number, and pressure, while also preserving the density recovery near the downstream region.
}
\label{fig:hardcase_centerline_compare}
\end{figure*}

The centerline comparison in Fig.~\ref{fig:hardcase_centerline_compare} provides a more interpretable view of the shock-localized improvement. Although all models recover the broad acceleration and recovery trends, the vanilla DeepONet and FNO baselines show a noticeable downstream shift of the shock-induced drop. The shock-aligned Fusion--DeepONet more accurately captures the shock station and the post-shock recovery, consistent with the quantitative shock-location metric in Table~\ref{tab:hardcase_model_comparison}.

The hard-case comparison confirms that the principal benefit of the proposed formulation is concentrated in the shock-sensitive region. This motivates two additional checks. First, we examine whether the selected Gaussian shock-envelope widths are robust to reasonable perturbations. Second, we verify that the reported behavior is not tied to a single optimizer choice. These checks are intended to support the final configuration rather than to introduce additional competing models.

\subsection{Sensitivity to shock-envelope widths}
\label{sec:rbf_sensitivity}

The shock-aligned trunk representation uses Gaussian envelopes to describe the neighborhood of the moving compression layer. Although the selected tuple \((2\Delta x,5\Delta x,8\Delta x)\) is physically motivated by the numerical shock thickness and its surrounding recovery region, the choice of envelope widths should not be treated as arbitrary. We therefore performed a sensitivity study in which the Gaussian tuple was varied while the network architecture, train/test split, and optimization protocol were kept fixed.

\begin{figure*}[t]
    \centering
    \includegraphics[width=0.78\textwidth]{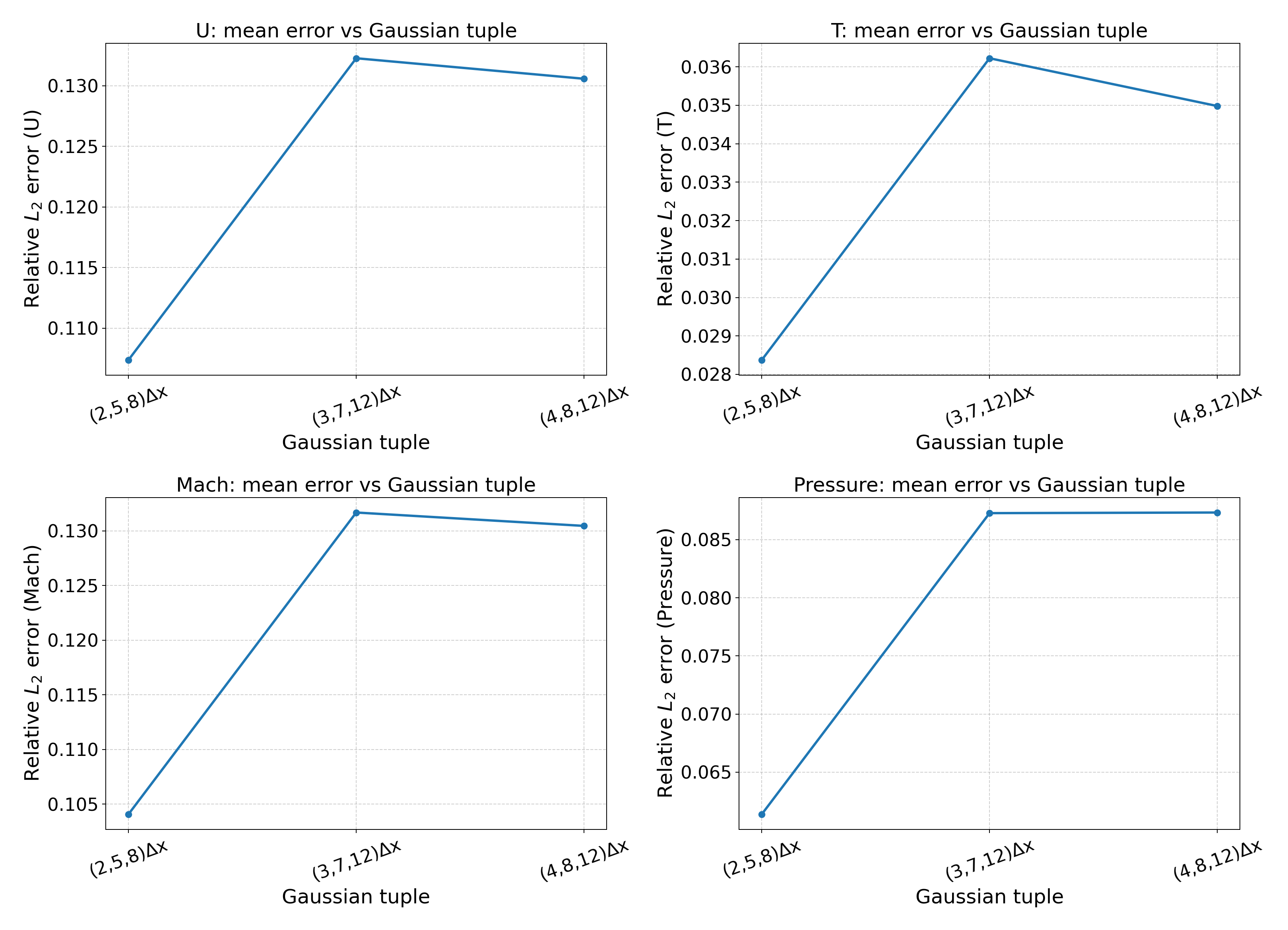}
    \caption{
Sensitivity of the shock-aligned Fusion--DeepONet to the Gaussian envelope widths used in the trunk features. The selected tuple \((2\Delta x,5\Delta x,8\Delta x)\) provides the best overall balance among the tested choices.
}
    \label{fig:rbf_mean_error}
\end{figure*}

Figure~\ref{fig:rbf_mean_error} summarizes the resulting mean relative \(L_2\) errors for the tested Gaussian tuples. The tuple \((2\Delta x,5\Delta x,8\Delta x)\) provides the best overall balance between localization of the shock core and representation of the adjacent recovery region. Wider envelopes remain stable but slightly reduce the ability of the trunk representation to localize the compression layer, whereas excessively broad features smear the shock neighborhood. The differences among reasonable tuples are moderate, indicating that the proposed shock-aligned representation is not overly sensitive to the exact envelope widths. Based on this study, \((2\Delta x,5\Delta x,8\Delta x)\) is used as the default setting in the final six-output model.

\subsection{Optimizer robustness}
\label{sec:optimizer_robustness}

The main training results reported above are obtained with the AdamW optimizer together with learning-rate reduction, gradient clipping, and early stopping. Since shock-sensitive neural-operator training can depend on the optimizer, we also tested a post-training full-batch limited-memory Broyden--Fletcher--Goldfarb--Shanno algorithm with bound constraints (L-BFGS-B) refinement stage for the selected shock-envelope tuple. The purpose of this test was not to introduce a separate model variant, but to verify that the reported behavior is not an artifact of a single first-order optimizer.

The refinement stage produced stable predictions and did not change the qualitative conclusions of the study. In particular, the dominant error mechanism remained the same: the largest discrepancies were localized near the compression layer, while the smoother pre-shock and post-shock regions were reconstructed consistently. This observation supports the interpretation that the proposed shock-aligned representation is responsible for the improved shock localization, rather than an optimizer-specific numerical artifact. For this reason, the final model is reported using the AdamW-trained configuration, while optimizer refinement is treated as a robustness check rather than a separate baseline.

\subsection{Nozzle with Throat Location Change}
\begin{figure}[t]
  \centering
  \includegraphics[width=0.95\linewidth]{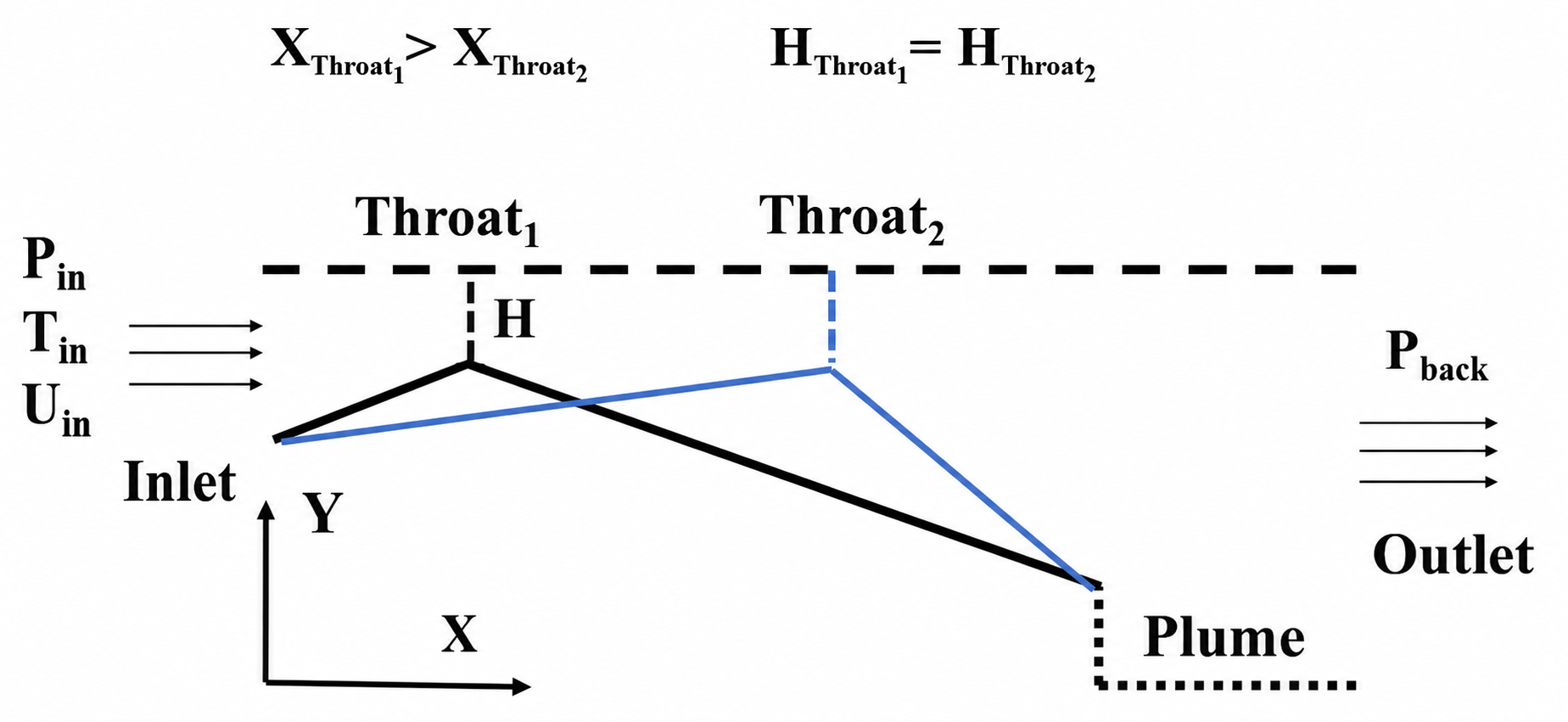}%
  \caption{Schematic of the micro-nozzle configuration with variable throat location and downstream plume region.}
  \label{fig:nozzle-schematic2}
\end{figure}

To quantify the sensitivity of the internal flow and plume to throat geometry, we vary only the axial throat position while keeping the throat height fixed, see Fig.~\ref{fig:nozzle-schematic2}. Specifically, the non-dimensional location is swept as
\(X_{\text{throat}}/L \in [0.10,\,0.55]\) at eight settings
(0.10, 0.15, 0.20, 0.25, 0.30, 0.35, 0.45, 0.55),
where \(L\) is the nozzle length. Inlet conditions \((P_{\text{in}}, T_{\text{in}}, U_{\text{in}})\) and outlet pressure \(P_{\mathrm{back}}\) are identical for all cases. One configuration (e.g., \(X_{\text{throat}}/L=0.30\)) is held out for testing; the rest are used for training/validation.

\begin{figure*}[t]
    \centering
    \includegraphics[width=0.95\textwidth]{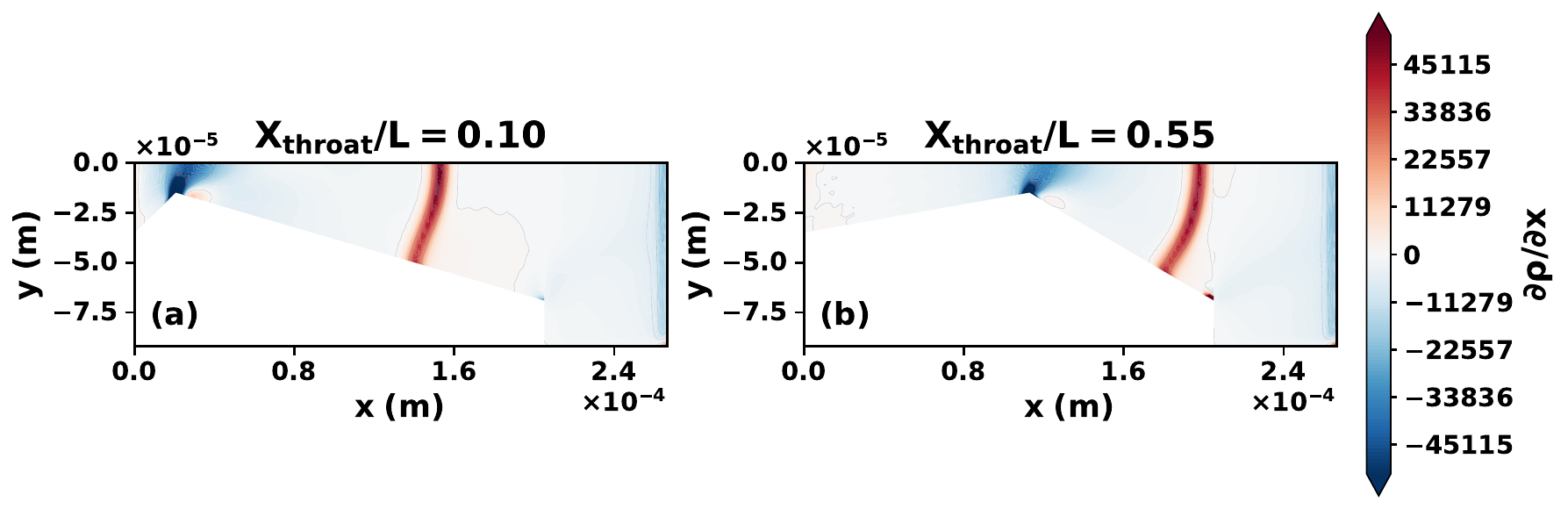}
    \caption{
    Streamwise density-gradient diagnostic for the two extreme throat-location cases. 
    Panels (a) and (b) show \(\partial \rho/\partial x\) for 
    \(X_{\mathrm{throat}}/L=0.10\) and \(X_{\mathrm{throat}}/L=0.55\), respectively. 
    The high-gradient band identifies the internal compression layer and illustrates the downstream displacement of the shock footprint as the throat is moved downstream.
    }
    \label{fig:throat_sweep}
\end{figure*}

To illustrate the sensitivity of the internal compression layer to the throat position, 
Fig.~\ref{fig:throat_sweep} compares the streamwise density-gradient field,
\(\partial \rho/\partial x\), for the two extreme throat locations in the sweep:
\(X_{\mathrm{throat}}/L=0.10\) and \(0.55\). 
The density-gradient diagnostic provides a sharper marker of the shock footprint than the primitive-variable contours. 
Although the inlet and outlet conditions are fixed, changing the throat position alters the effective expansion length and the downstream adjustment of the flow. 
As a result, the dominant high-gradient band associated with the compression layer shifts downstream as the throat is moved from \(X_{\mathrm{throat}}/L=0.10\) to \(0.55\). 
These two cases bound the observed range of shock locations over the throat-position interval considered.

\subsection{Generalization to an unseen throat location}

To further examine geometric generalization, we consider a throat-location case that was entirely excluded from training. In this test, the throat is located at
\[
X_{\mathrm{throat}}/L=0.30,
\]
whereas the training set contains the neighboring throat locations
\[
X_{\mathrm{throat}}/L=\{0.10,0.15,0.20,0.25,0.35,0.45,0.55\}.
\]
Thus, this case provides a direct interpolation test with respect to the axial throat position.

\begin{figure}[t]
  \centering
  \includegraphics[width=\textwidth]{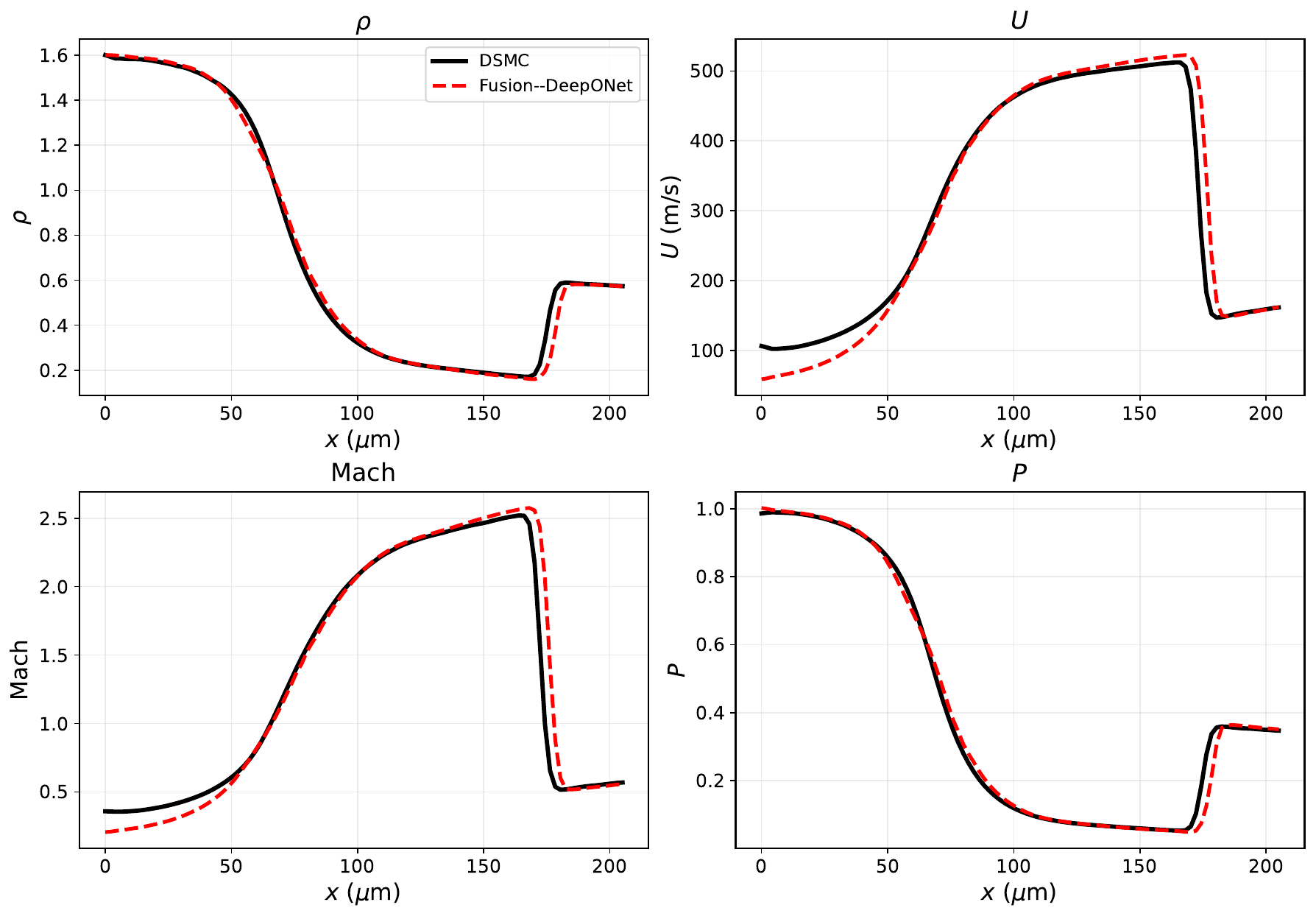}
  \caption{
  Centerline comparison between the DSMC reference solution and the Fusion--DeepONet prediction for the held-out throat-location case
  \(X_{\mathrm{throat}}/L=0.30\). The comparison includes density, streamwise velocity, Mach number, and pressure.
  }
  \label{fig:throat_test030_centerline}
\end{figure}

Figure~\ref{fig:throat_test030_centerline} compares the DSMC centerline solution with the Fusion--DeepONet prediction for density, streamwise velocity, Mach number, and pressure. The model captures the smooth upstream acceleration, the expansion through the throat, the low-pressure high-Mach region, and the post-shock recovery. The density and pressure profiles show the expected monotonic decrease through the converging--diverging section followed by a sharp recovery across the internal shock. The corresponding increase of \(U\) and Mach number upstream of the shock is also reproduced.

The main pointwise error mechanism is again localized near the compression
layer. The DSMC profile contains a sharp downstream transition, while the
neural prediction produces a slightly smoother transition. This is consistent
with the sensitivity of shock-dominated profiles to small streamwise shifts in
the compression-layer location. Importantly, the upstream acceleration, the
low-pressure high-Mach region, and the post-shock thermodynamic recovery are
all reproduced with the correct trend and magnitude.

The quantitative centerline errors are summarized in Table~\ref{tab:throat_test030_min_error}. 
For the held-out throat-location case \(X_{\mathrm{throat}}/L=0.30\), the model achieves relative \(L_2\) errors of
\(4.07\%\), \(9.78\%\), \(10.85\%\), and \(4.82\%\) for density, streamwise velocity, Mach number, and pressure, respectively.
The pressure and density predictions are the most accurate among the four reported quantities, while the larger errors in \(U\) and Mach number are mainly associated with the slight shift and smoothing of the internal shock location.
Overall, these results show that the Fusion--DeepONet prediction captures the dominant centerline flow evolution for an unseen throat position, with the main discrepancy confined to the shock region.

\begin{table}[t]
\centering
\caption{
Minimum centerline error metrics obtained for the held-out throat-location case
\(X_{\mathrm{throat}}/L=0.30\).
}
\label{tab:throat_test030_min_error}
\begin{tabular}{lc}
\toprule
Variable & Relative \(L_2\) error \\
\midrule
Density  & 0.0407  \\
\(U\)    & 0.0978  \\
Mach     & 0.1085 \\
Pressure & 0.0482  \\
\bottomrule
\end{tabular}
\end{table}

\clearpage
\section{Concluding Remarks}
\label{sec:conclusion}

This study developed a shock-aligned Fusion--DeepONet surrogate for DSMC-generated rarefied micro-nozzle flows with moving internal compression layers. The central idea is to introduce the dominant shock topology directly into the trunk representation through a training-derived signed distance to the shock, a smooth pre-/post-shock indicator, multi-scale Gaussian envelopes, and a zone marker. The resulting model remains data-driven, since no governing-equation residuals or jump conditions are imposed, but it incorporates physically meaningful alignment information that reduces the burden of learning a moving shock in Cartesian coordinates.

The six-output model was evaluated on held-out back-pressure cases of \(P_{\mathrm{back}}=16\), 25, and 30 kPa. Across these cases, the surrogate predicted density, temperature, and pressure with relatively low global errors, while the larger errors in \(U\), \(V\), and Mach number were mainly localized near the shock-sensitive regions. This behavior is physically consistent: small offsets in the compression-layer location produce amplified pointwise errors in velocity and Mach number, even when the global flow topology is correctly reconstructed. Centerline and contour comparisons confirmed that the model captures the main acceleration, shock-induced drop, and downstream recovery trends.

The exit-plane engineering diagnostics further demonstrated that the surrogate preserves quantities relevant to nozzle-performance assessment. The total thrust-equivalent exit-plane flux was predicted within \(1.6\%\) for all three held-out back-pressure cases, and the mass flow rate was recovered within \(1\%\) for the 25 and 30 kPa cases. For the most challenging 16 kPa condition, the integrated exit-plane response
remains accurate despite the expected local sensitivity of pointwise
shock-region errors. This confirms that the surrogate preserves the
engineering-relevant nozzle response while localizing the remaining field
errors to the compression-layer neighborhood.

A key finding of this work is that the DSMC-resolved internal compression layer admits a compact shock-centered representation. In the physical coordinate \(x\), the leading POD mode of the density profiles captures only \(83.33\%\) of the fluctuation energy. After registration with the jump-scaled coordinate \(\xi_j=(x-x_s)/\delta_j\), this value increases to \(98.33\%\). This confirms that much of the apparent parametric complexity is caused by translation and finite-thickness scaling of the compression layer, rather than by a fully high-dimensional deformation of the entire flow field. The shock-aligned trunk representation exploits this reduced structure directly, which explains why the proposed model improves shock-window and gradient-weighted errors more strongly than global errors.

The hard-case comparison at \(P_{\mathrm{back}}=16~\mathrm{kPa}\) showed the clearest benefit of the proposed shock-aligned formulation. The shock-aligned Fusion--DeepONet achieved a global mean error of \(3.15\%\), a shock-window mean error of \(4.51\%\), and a gradient-weighted mean error of \(3.53\%\), outperforming the vanilla MLP, vanilla DeepONet, and FNO baselines. In particular, the proposed model reduced the shock-window error relative to the vanilla DeepONet and FNO baselines and eliminated the measured shock-location offset based on the velocity-gradient peak. These results support the conclusion that the improvement is not merely due to network capacity, but to the shock-aligned trunk representation.

The geometric generalization test with variable throat location provided an additional assessment of the framework beyond back-pressure interpolation. For the held-out throat position \(X_{\mathrm{throat}}/L=0.30\), the model achieved centerline relative \(L_2\) errors of \(4.07\%\), \(9.78\%\), \(10.85\%\), and \(4.82\%\) for density, streamwise velocity, Mach number, and pressure, respectively. The main discrepancy again occurred near the shock, while the upstream acceleration and downstream thermodynamic recovery were recovered well. This result suggests that the same operator-learning strategy can be extended from operating-condition variation to geometry-driven variation when an appropriate alignment coordinate is available.
While the present study focuses on a single planar argon micro-nozzle family, the central finding---that a rarefied flow containing a localized, finite-thickness compression layer admits a compact reduced representation once the dominant translational motion and thickness variation of that layer are removed---is expected to be relevant to a broader class of rarefied internal and near-field flows that exhibit moving compression or shock structures, including micro-step flows, shock--boundary-layer interactions in MEMS devices, and certain plume impingement configurations.
The scope of the present study is intentionally restricted to a single planar
argon micro-nozzle family with fixed inlet conditions and back-pressure-driven
motion of the internal compression layer, together with one held-out
throat-location test. Therefore, the results should not be interpreted as a
universal scaling law for all rarefied shocks or nozzle geometries. The more
general implication is that rarefied flows containing a localized,
finite-thickness compression layer may admit a compact representation when the
dominant motion and thickness variation of that layer are removed. Future work
should test this hypothesis across different gases, wall-scattering models,
inlet Knudsen numbers, nozzle contours, and stronger geometry variations.

The scope of the present study is intentionally restricted to a single planar
argon micro-nozzle family with fixed inlet conditions, specular wall reflection,
and back-pressure-driven motion of the internal compression layer, together
with one held-out throat-location test. Therefore, the results should not be
interpreted as a universal scaling law for all rarefied shocks, gas--surface
models, or nozzle geometries. The broader implication is more specific: when a
rarefied flow contains a localized, finite-thickness, high-gradient structure
whose dominant variability is translation and thickness modulation, a
shock-centered or layer-centered registration can substantially reduce the
apparent parametric dimension of the DSMC fields. This strategy is therefore
expected to be relevant beyond the present micro-nozzle, including rarefied
micro-step flows with moving separation or recirculation regions,
shock--boundary-layer interactions in MEMS-scale devices, confined micro-jet
expansions, and internal rarefied flows in which compression, expansion, or
Knudsen-layer structures move with operating condition or geometry. In such
problems, the same principle can be used to construct physics-aware neural
operators: first identify the dominant kinetic layer through DSMC-based
gradient or gradient-length diagnostics, then encode its location, signed
distance, and finite spatial support as inductive-bias features. Future work
should test this hypothesis across different gases, accommodation models,
diffuse and mixed gas--surface reflections, inlet Knudsen numbers, nozzle
contours, and stronger geometry variations.

\subsection*{Data availability}
The scripts used to compute the shock-centered diagnostics, POD spectra,
kinetic-scale tables, and figure data will be available at
\url{https://github.com/Ehsan-Roohi/} upon acceptance of the paper.

\appendix
\section{Burgers Benchmark}
\label{app:burgers}

\subsection{Canonical Validation on Burgers’ Equation}
\begin{figure*}[t]
  \centering
  \includegraphics[width=1.05\textwidth]{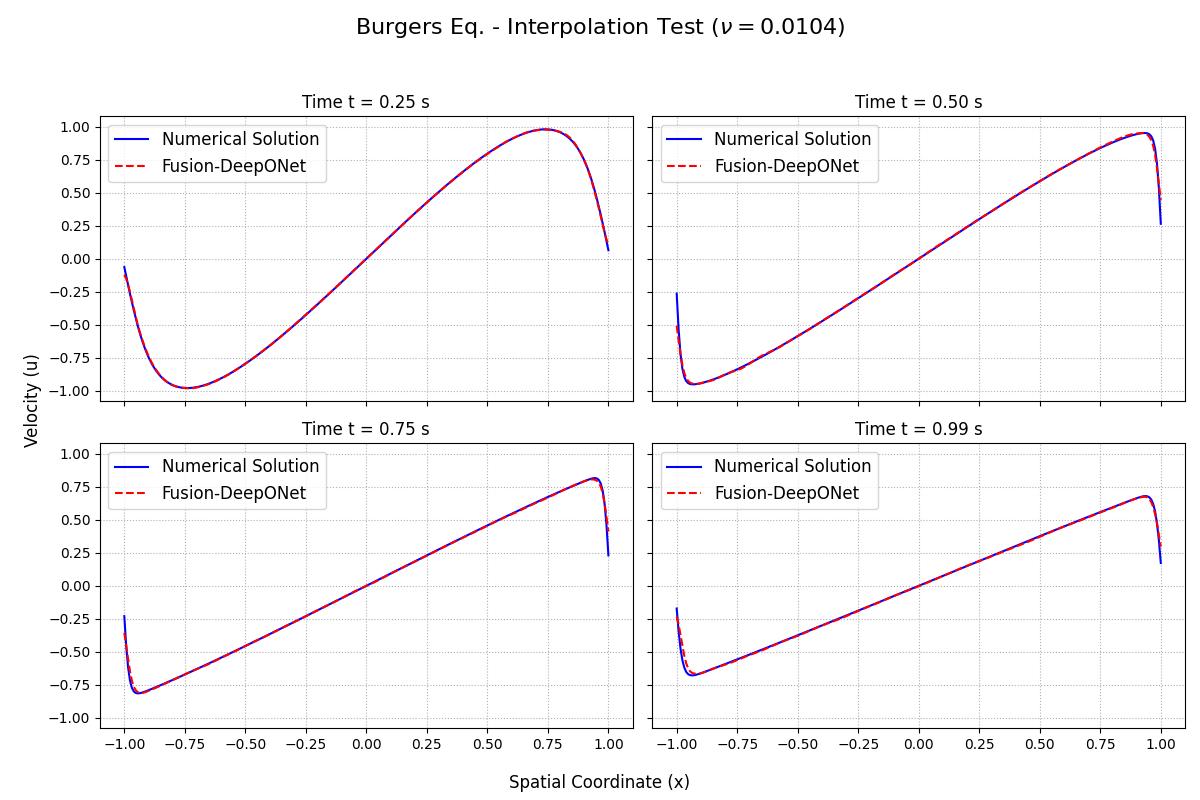}
  \vspace{-0.5em}
  \caption{Interpolation test of the Burgers' equation.}
  \label{fig:burgers-interp}
\end{figure*}

\begin{figure*}[t]
  \centering
  \includegraphics[width=1.05\textwidth]{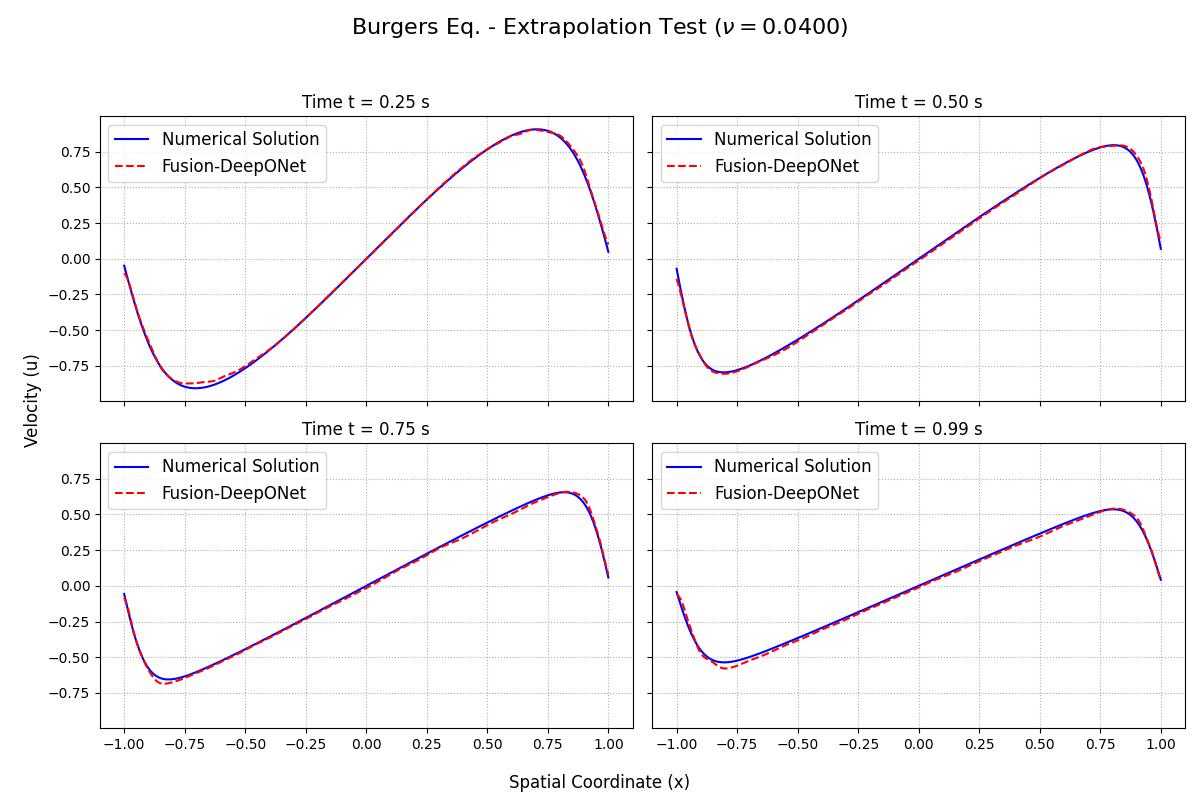}
  \vspace{-0.5em}
  \caption{
Near-boundary parameter test of Burgers' equation.}
  \label{fig:burgers-extra}
\end{figure*}

We use the viscous Burgers equation as a controlled representation check before
applying the method to the DSMC nozzle data. The benchmark tests whether the
shock-aligned feature construction can represent a moving localized gradient in
a simple setting. Figure~\ref{fig:burgers-interp} reports an interpolation case,
while Fig.~\ref{fig:burgers-extra} shows a near-boundary parameter test. The
model reproduces the main phase, amplitude, and steep-front behavior, with
relative \(L_2\) errors of 2.36\% and 2.69\%, respectively.
Overall, the Burgers benchmark is used only as a controlled representation check. The main evidence for the usefulness and limitations of the proposed surrogate is provided by the DSMC micro-nozzle tests in the main text.

\section{Consistency of density- and velocity-gradient shock stations}
\label{app:xs_consistency}

\begin{figure}[t]
\centering
\includegraphics[width=0.62\textwidth]{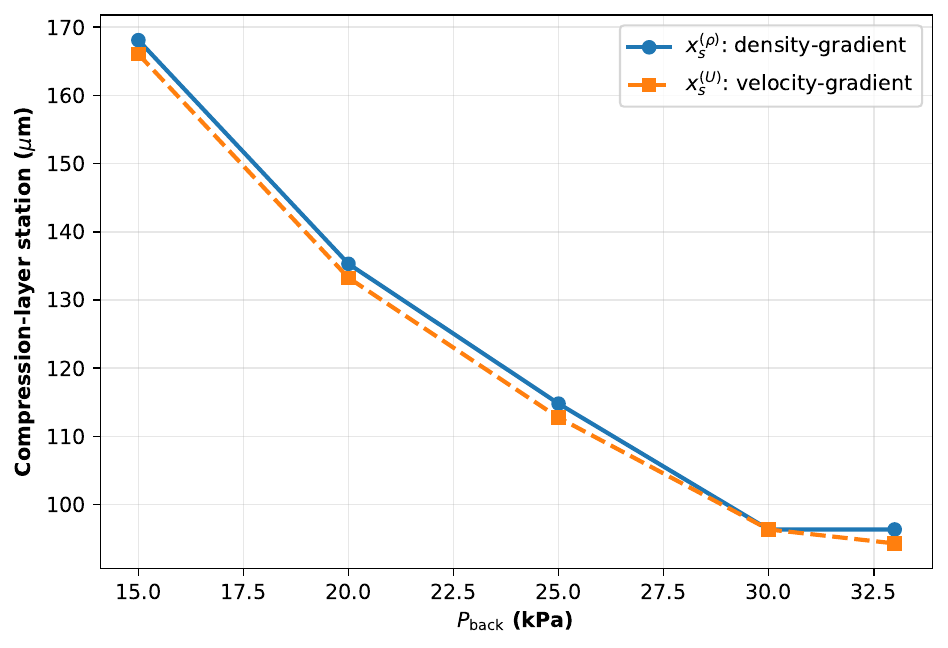}
\caption{
Comparison between the density-gradient and velocity-gradient definitions of
the compression-layer station. The two diagnostics identify nearly the same
moving compression region over the representative back-pressure range.
}
\label{fig:xs_consistency_plot}
\end{figure}

The shock-centered reduced-order analysis uses the density-gradient maximum to
define the compression-layer station, whereas the neural-operator alignment map
is calibrated from the streamwise velocity-gradient maximum. To verify that these two definitions identify the same internal compression
region, we compare the two station estimates for the representative
back-pressure cases used in the reduced-order diagnostics in
Table~\ref{tab:xs_consistency} and Fig.~\ref{fig:xs_consistency_plot}.

\begin{table}[t]
\centering
\caption{
Consistency between density-gradient and velocity-gradient compression-layer
stations for representative DSMC cases.
}
\label{tab:xs_consistency}
\begin{tabular}{ccccc}
\toprule
\(P_{\mathrm{back}}\) &
\(x_s^{(\rho)}\) &
\(x_s^{(U)}\) &
\(\Delta x_s\) &
\(\Delta x_s/\Delta x\) \\
(kPa) & (\(\mu\)m) & (\(\mu\)m) & (\(\mu\)m) & -- \\
\midrule
15 & 168.10 & 166.05 & 2.05 & 1.00 \\
20 & 135.30 & 133.25 & 2.05 & 1.00 \\
25 & 114.80 & 112.75 & 2.05 & 1.00 \\
30 & 96.35  & 96.35  & 0.00 & 0.00 \\
33 & 96.35  & 94.30  & 2.05 & 1.00 \\
\bottomrule
\end{tabular}
\end{table}

Across the representative cases, the mean difference between
\(x_s^{(\rho)}\) and \(x_s^{(U)}\) is \(1.64~\mu\mathrm{m}\), corresponding to
\(0.80\) grid spacings, while the maximum difference is \(2.05~\mu\mathrm{m}\),
corresponding to \(1.00\) grid spacing. Thus, the density-gradient station used
for physical diagnostics and the velocity-gradient station used for the
training-calibrated alignment map are consistent at the resolution of the DSMC
fields.

\section{Additional Contours for Transverse Velocity and Temperature}
\label{app:vt_contours}

\begin{landscape}
\begin{figure}[p]
\centering
\includegraphics[
width=1.02\linewidth,
trim=0 0 0 1.0cm,
clip
]{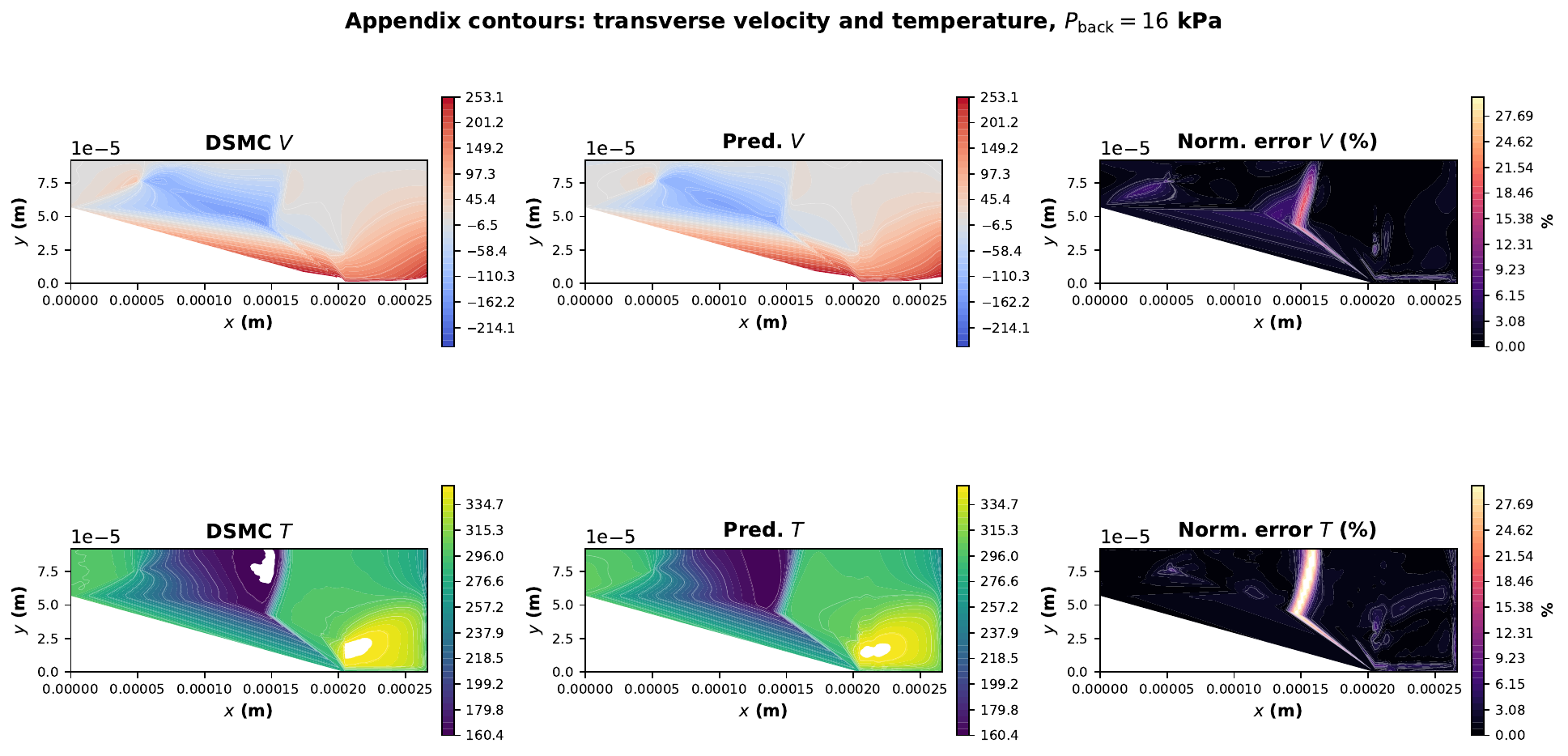}
\caption{
Additional transverse-velocity and temperature contour comparisons for \(P_{\mathrm{back}}=16~\mathrm{kPa}\). The surrogate reproduces the main spatial organization of both \(V\) and \(T\), while the largest normalized errors remain localized near the compression layer and high-gradient regions.
}
\label{fig:appendix_vt_pr16}
\end{figure}
\end{landscape}

\begin{landscape}
\begin{figure}[p]
\centering
\includegraphics[
width=1.02\linewidth,
trim=0 0 0 1.0cm,
clip
]{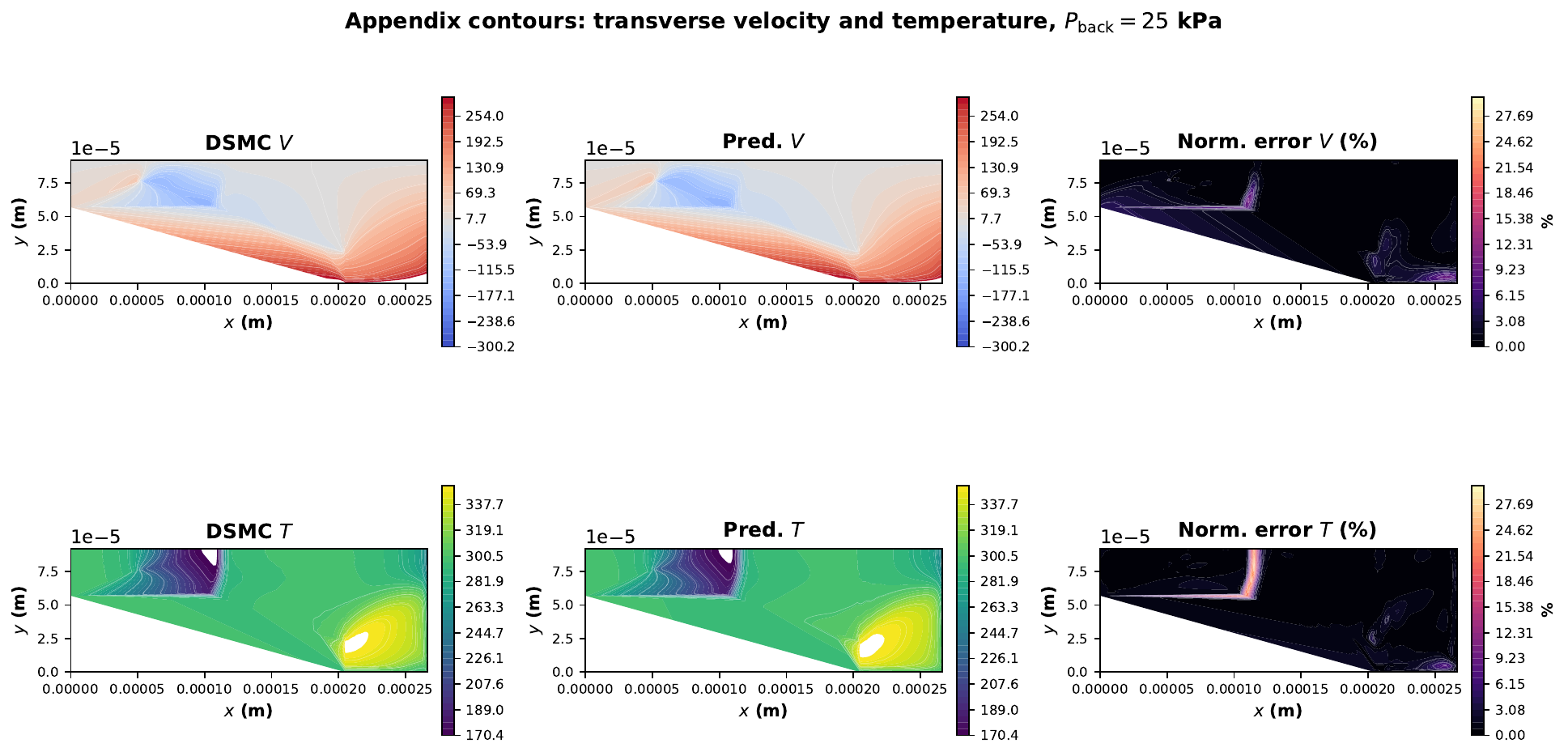}
\caption{
Additional transverse-velocity and temperature contour comparisons for \(P_{\mathrm{back}}=25~\mathrm{kPa}\). The DSMC and surrogate fields show close qualitative agreement, with residual discrepancies concentrated near the shock-adjacent region.
}
\label{fig:appendix_vt_pr25}
\end{figure}
\end{landscape}

\begin{landscape}
\begin{figure}[p]
\centering
\includegraphics[
width=1.02\linewidth,
trim=0 0 0 1.0cm,
clip
]{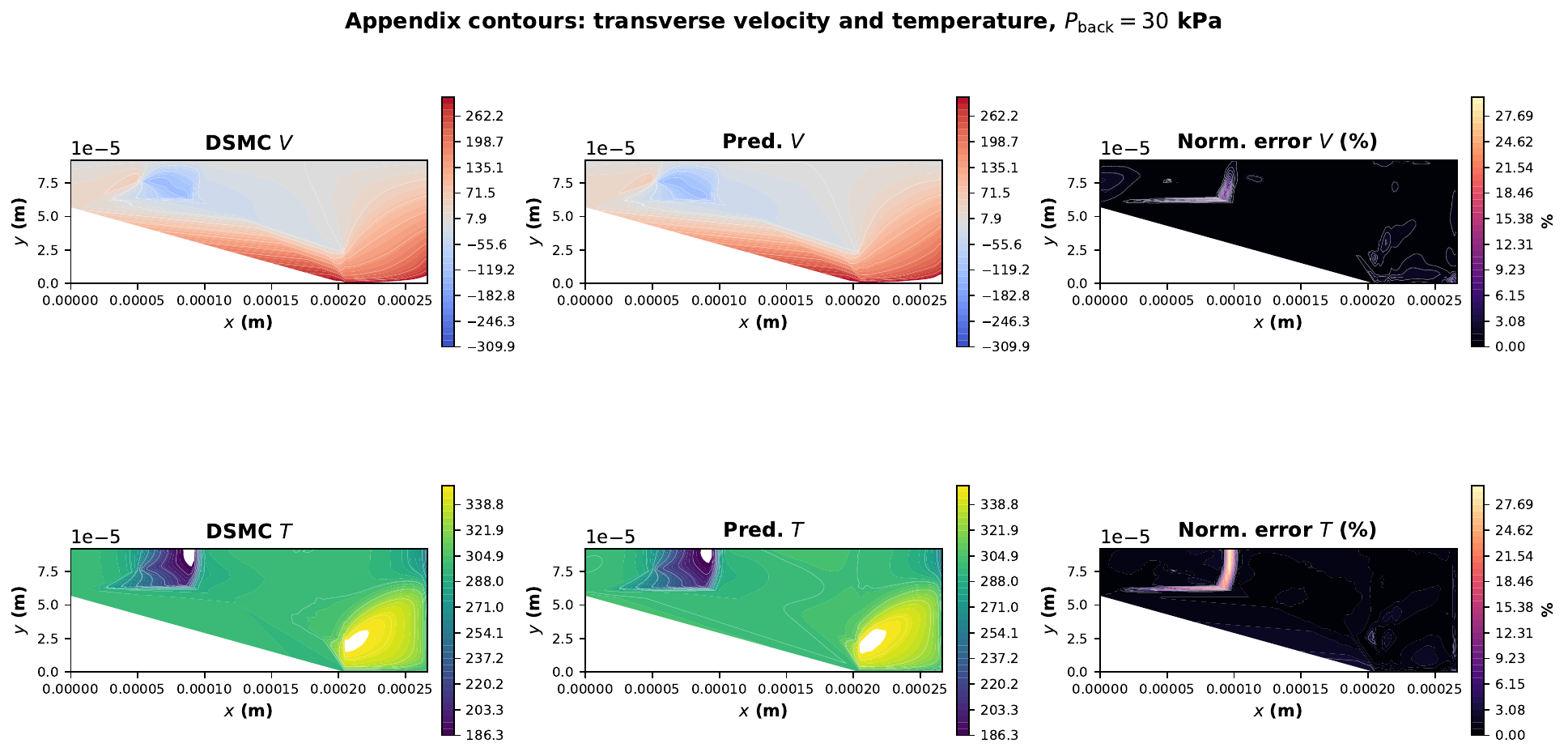}
\caption{
Additional transverse-velocity and temperature contour comparisons for \(P_{\mathrm{back}}=30~\mathrm{kPa}\). The results confirm that the surrogate also reconstructs the auxiliary transverse-velocity and temperature fields for the high-back-pressure held-out case.
}
\label{fig:appendix_vt_pr30}
\end{figure}
\end{landscape}

The main text reports contour comparisons for density, streamwise velocity, Mach number, and pressure. These variables are selected because they most clearly show the shock displacement, acceleration, and pressure-recovery behavior. For completeness, this appendix reports the two additional predicted fields, transverse velocity \(V\) and temperature \(T\), for the held-out back-pressure cases. Each row compares the DSMC reference field, the surrogate prediction, and the normalized error diagnostic,
\[
E_q=\frac{|\hat q-q|}{q_{99}-q_1},
\]
where \(q_1\) and \(q_{99}\) are the 1st and 99th percentiles of the DSMC reference field for each variable. The color scale is clipped for visualization so that localized shock-adjacent errors do not obscure the global field agreement. The \(V\) and \(T\) contour comparisons are shown in
Figs.~\ref{fig:appendix_vt_pr16}, \ref{fig:appendix_vt_pr25}, and \ref{fig:appendix_vt_pr30}.

\clearpage

\clearpage

\clearpage

\subsection{Computational cost}
\label{sec:computational_cost}

The main computational advantage of the proposed surrogate appears in repeated
parametric evaluation after the DSMC database has been generated. Therefore, the relevant comparison is the
marginal cost of evaluating the trained surrogate relative to the cost of
generating an additional statistically converged DSMC solution. For the present
micro-nozzle cases, a typical DSMC simulation requires approximately
\(10\)--\(15\) hours of CPU wall-clock time, depending on the back pressure and
boundary-condition details. In contrast, training and inference of the
Fusion--DeepONet surrogate require less than \(30\) minutes on a high-end GPU
for the set of configurations considered here, as summarized in
Table~\ref{tab:computational_cost}.

Because the DSMC solver and the neural surrogate use different hardware
architectures and perform fundamentally different tasks, this comparison should
not be interpreted as a hardware-independent algorithmic speed-up. Rather, it
indicates the practical reduction in marginal evaluation cost once the DSMC
training database is available. 
The reported surrogate time includes the one-time training stage and subsequent
inference for the configurations considered here; the marginal inference cost
after training is substantially smaller. Thus, the comparison is conservative
with respect to repeated-query use.

\begin{table}[t]
\centering
\caption{Indicative computational cost comparison.}
\label{tab:computational_cost}
\begin{tabular}{lcc}
\toprule
Method & Hardware & Wall-clock time \\
\midrule
DSMC reference solver & CPU & \(10\)--\(15\) h per case \\
Fusion--DeepONet train+infer & GPU & \(<30\) min total \\
\bottomrule
\end{tabular}
\end{table}

\clearpage
\bibliographystyle{unsrtnat} 
\bibliography{references}  

\end{document}